\documentclass[12pt]{article}
\usepackage{epsf}

\usepackage{epsfig,graphics}
\usepackage {graphicx}
\usepackage {epsfig}
\usepackage {subfigure}
\usepackage {tabularx} 
\usepackage{rotate}	
\usepackage{slashed}

\newcommand{\bmat}{\left(\begin{array}}
\newcommand{\emat}{\end{array}\right)}

\def\yzero{\smash{\hbox{$y\kern-4pt\raise1pt\hbox{${}^\circ$}$}}}

\def\beq{\begin{equation}}
\def\eeq{\end{equation}}
\def\beqa{\begin{eqnarray}}
\def\eeqa{\end{eqnarray}}

\def\-{\hphantom{-}}

\def\s2{\frac{1}{\sqrt2}}

\def\beq{\begin{equation}}
\def\eeq{\end{equation}}
\def\beqa{\begin{eqnarray}}
\def\eeqa{\end{eqnarray}}

\def\IF{\relax{\rm I\kern-.18em F}}
\def\II{\relax{\rm I\kern-.18em I}}
\def\IP{\relax{\rm I\kern-.18em P}}
\def\IC{\relax\hbox{\kern.25em$\inbar\kern-.3em{\rm C}$}}
\def\IR{\relax{\rm I\kern-.18em R}}

\def\Dsl{\,\raise.15ex\hbox{/}\mkern-13.5mu D} 
\def\IZ{Z\kern-.4em  Z}

\def\crosssec{\sigma_{\tilde{\chi}_1^0-p}}
\def\lsim{\raise0.3ex\hbox{$\;<$\kern-0.75em\raise-1.1ex\hbox{$\sim\;$}}}
\def\gsim{\raise0.3ex\hbox{$\;>$\kern-0.75em\raise-1.1ex\hbox{$\sim\;$}}}

\def\met{\slash\hspace*{-1.5ex}E_T}

\catcode`\@=11   
\newdimen\@rotdimen
\newbox\@rotbox  

\def\@vspec#1{\special{ps:#1}}
\def\@rotstart#1{\@vspec{gsave currentpoint currentpoint translate
   #1 neg exch neg exch translate}}
\def\@rotfinish{\@vspec{currentpoint grestore moveto}}
\def\@rotr#1{\@rotdimen=\ht#1\advance\@rotdimen by\dp#1%
   \hbox to\@rotdimen{\hskip\ht#1\vbox to\wd#1{\@rotstart{90 rotate}%
   \box#1\vss}\hss}\@rotfinish}
\def\@rotl#1{\@rotdimen=\ht#1\advance\@rotdimen by\dp#1%
   \hbox to\@rotdimen{\vbox to\wd#1{\vskip\wd#1\@rotstart{270 rotate}%
   \box#1\vss}\hss}\@rotfinish}%
\def\@rotu#1{\@rotdimen=\ht#1\advance\@rotdimen by\dp#1%
   \hbox to\wd#1{\hskip\wd#1\vbox to\@rotdimen{\vskip\@rotdimen
   \@rotstart{-1 dup scale}\box#1\vss}\hss}\@rotfinish}%
\def\@rotf#1{\hbox to\wd#1{\hskip\wd#1\@rotstart{-1 1 scale}%
   \box#1\hss}\@rotfinish}%
\def\rotate{\@ifnextchar[{\@rotate}{\@rotate[l]}}
\def\@rotate[#1]#2{\setbox\@rotbox=\hbox{#2}\@nameuse{@rot#1}\@rotbox}

\catcode`\@=12

\topmargin
-1.5cm
\textwidth
15.5cm
\textheight
23.5cm
\oddsidemargin
0.7cm
\evensidemargin
0.7cm

\begin{document}

\makeatletter
\@addtoreset{equation}{section}
\makeatother
\renewcommand{\theequation}{\thesection.\arabic{equation}}
\pagestyle{empty}
\rightline{ IFT-UAM/CSIC-12-12}
\rightline{ FTUAM-12-83}
\vspace{0.1cm}
\begin{center}
\LARGE{A $119-125$ GeV Higgs from \\
a string derived  slice of the CMSSM 
  \\[5mm]}
 \large{L. Aparicio, D.G. Cerde\~no and L.E. Ib\'a\~nez \\[6mm]}
\small{
 Departamento de F\'{\i}sica Te\'orica
and Instituto de F\'{\i}sica Te\'orica UAM/CSIC,\\[-0.3em]
Universidad Aut\'onoma de Madrid,
Cantoblanco, 28049 Madrid, Spain 
\\[8mm]}
\small{\bf Abstract} \\[5mm]
\end{center}
\begin{center}
\begin{minipage}[h]{15.0cm}
The recent experimental hints for a relatively heavy Higgs with a mass in the range  $119-125$~GeV favour supersymmetric 
scenarios with a large mixing in the stop mass matrix. It has been shown that this is possible in the constrained Minimal Supersymmetric Standard Model (CMSSM), but only for a very specific relation between the trilinear parameter and the soft scalar mass, 
favouring $A\approx-2m$ for a relatively light spectrum, and sizable values of $\tan\beta$. 
We describe  here a string-derived  scheme  in which the first condition is automatic  and the second
 arises as a consequence of imposing radiative EW symmetry breaking and viable neutralino dark matter in agreement with WMAP constraints. 
More specifically, we consider modulus dominated SUSY-breaking in Type II string compactifications and show that it leads   to a very predictive CMSSM-like scheme, with small departures due to background fluxes. Imposing the above constraints leaves only one free parameter, which corresponds to an overall scale. We show that in this construction $A=-3/\sqrt{2}m\simeq -2m$ and in the allowed parameter space $\tan\beta \simeq 38-41$,  leading to $119$~GeV $< m_h< 125$~GeV. 
The recent LHCb results on BR$(B_s\to\mu^+\mu^-)$ further constrain this range, leaving only the region with $m_h\sim125.$~GeV. 
We determine the detectability of this model and show that it could start being probed by the LHC at 7(8)~TeV with a luminosity of 5(2)~fb$^{-1}$, and the whole parameter space would be accessible for $14$~TeV and 25~fb$^{-1}$. 
Furthermore, this scenario can host a long-lived stau with the right properties to lead to catalyzed BBN.
We finally argue  that anthropic arguments could favour the highest value for the Higgs mass that is compatible with neutralino dark matter, i.e., $m_h\sim125$~GeV.

\end{minipage}
\end{center}
\newpage
\setcounter{page}{1}
\pagestyle{plain}
\renewcommand{\thefootnote}{\arabic{footnote}}
\setcounter{footnote}{0}

\section{Introduction}

With the advent of the Large Hadron Collider (LHC) Particle Physics is entering into a new era in which a wealth of 
theoretical models, scenarios and ideas are being tested. One of the most prominent ideas
 beyond the Standard Model (SM)  is low energy supersymmetry (SUSY) and its simplest implementation, the Minimal Supersymmetric Standard Model (MSSM).
 Although at the moment no sign of supersymmetric particles has been seen, there is
 at least one recent LHC result which points in the direction of supersymmetry.
 The 2011 run of LHC has restricted the most likely 
 range for the  Higgs particle mass to be  $115.5-131$~GeV (ATLAS) \cite{ATLASHiggs} and $114.5-127$ GeV (CMS) \cite{CMSHiggs}.
 In addition, there are  hints observed by both CMS and ATLAS of an excess of  events that
 might correspond to $\gamma \gamma$, $ZZ^*\rightarrow 4l$ and $WW^*\rightarrow 2l$ decays
 of a Higgs particle with a mass in a range close to $125$ GeV.  
 Interestingly, such values for the Higgs mass are consistent with the expected range $<130$ GeV for the lightest
 Higgs in the MSSM.

  Although in qualitative agreement with MSSM expectations, the hints of a $125$~GeV Higgs are
     slightly uncomfortable for  models like the Constrained Minimal Supersymmetric Standard Model (CMSSM),
      in which the complete SUSY spectra is determined 
  in terms of a few universal soft supersymmetry-breaking parameters $M,\,m,\,A,\,B,\,$ and $\mu$ \cite{CMSSM}. 
  Indeed lighter Higgs masses of order
  $110-115$~GeV are generic  in the CMSSM parameter space. In order to get values as large as
  $125$~GeV one needs to have heavy stops with a sizable LR-mixing and large values of $\tan\beta$, leading typically to a very
  heavy  SUSY spectrum. In fact it has been noted 
   \cite{Baer:2011ab,Hall:2011aa,Arbey:2011ab,Akula:2011aa} (see also \cite{Gogoladze:2011aa,Carena:2011aa})
  that the areas of the CMSSM parameter space compatible 
  with $125$ GeV Higgs show a very strong preference for the region with $A\simeq -2m$ if the SUSY spectrum is not
  to be very heavy
  \footnote{Note in passing that a $125$ GeV Higgs is difficult to accommodate in the simplest gauge mediation scenarios
  since $A=0$ in these schemes, see Refs.\,\cite{Arbey:2011ab,Draper:2011aa,Evans:2012hg}.}.  But why 
  should nature be centered in that peculiar corner of parameter space?

  A possible explanation for relations among soft terms like, e.g., $A$ and $m$ requires going beyond the 
  general assumptions underlying the CMSSM scheme and being more specific about the origin of
  SUSY breaking. The CMSSM boundary conditions are obtained in supergravity mediation schemes with
  unification (GUT-like) constraints and universal kinetic terms for all the SM matter fields. In order to get 
  relations among the  $M,\,m,\,A,\,M,\,\mu$ parameters one needs very 
  specific classes of low energy $N=1$ supergravity models. It is here where string unification models
  arising from specific classes of string compactifications may be useful.

  Indeed, in low-energy supergravity models coming from string compactifications the gauge kinetic functions 
  as well as the kinetic (Kahler metrics) terms of the SM fields are not arbitrary and depend on the moduli 
  of the corresponding string compactification.  If the auxiliary fields of the moduli are the source
  of SUSY-breaking, specific relations among the different soft terms are obtained.
  These have been worked out for heterotic vacua  
  \cite{Brignole:1993dj,Cvetic:1991qm,il,Kaplunovsky:1993rd} (see e.g.  Ref.\,\cite{bim} for a review and further references) 
  and generalized for the more recent case of
  Type II orientifold compactifications  
  \cite{imr,fluxed,aci}. See also Ref.\,\cite{othersoft} and references therein 
  for explicit SUSY-breaking models in  Type II orientifolds.

In the last decade there has been important progress in the construction of semirealistic Type II
string vacua. With the advent of the D-brane techniques  it has been possible to construct
Type II   string orientifold configurations of branes yielding a massless spectrum close to that of the MSSM (see Ref.\,\cite{book}
for a review).  A particularly successful scheme is the one based on Type IIB orientifolds with 
the SM fields residing on intersecting  7-branes and their non-perturbative generalization, F-theory.
One of the attractive aspects of this large class of compactifications is that it is well understood how
the presence of antisymmetric field fluxes and possibly non-perturbative effects  can give rise to a complete 
fixing of the moduli of the compactification \cite{kklt} (for reviews see 
Refs.\,\cite{Denef:2008wq,Douglas:2006es,book}).
 In addition, the large number of possible fluxes allows to 
fine-tune the  vacuum energy to a small but positive value, in a way compatible with 
a non-vanishing positive cosmological constant.

Besides fixing the moduli, such fluxes in general  give rise
to soft SUSY breaking terms for the MSSM fields in semirealistic compactifications
\cite{softfromflux}.
 In particular, it has 
been found that certain ISD (imaginary self-dual) fluxes correspond to the presence of Kahler modulus 
dominated SUSY-breaking, providing an explicit realization of gravity mediation SUSY-breaking in string
theory. Such type of fluxes are important since it has also been shown that they are consistent with
the classical equations of motion of Type IIB orientifolds \cite{Giddings:2001yu}.

\begin{figure}[h!]
\hspace*{-0.6cm}
\centering
\includegraphics[width=11.cm, angle=0]{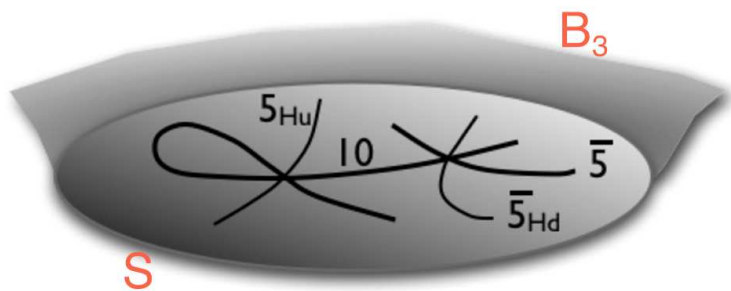}
\caption{General structure of a local  F-theory $SU(5)$ GUT. The GUT group lives on 7-branes whose 4
extra dimensions beyond Minkowski wrap a 4-cycle $S$. This $S$ manifold is inside a 3 complex dimensional 
manifold $B_3$ where the 6 extra dimensions are compactified. The gauge bosons live in the bulk of $S$ whereas
quarks,  leptons, and Higgsses are localized in complex curves inside $S$. These matter curves ($10$ and ${\bar 5}$ in
the figure) correspond to the intersection of 
the 7-branes wrapping $S$ with other $U(1)$  7-branes (not depicted in the figure).  There is one matter curve for each $SU(5)$ rep. and at the
intersection of matter curves with Higgs curves $H_u, H_d$ Yukawa couplings develop (figure taken  from Ref.\,\cite{book}).
}
\label{ftheoryguts}
\end{figure}

In Ref.\,\cite{aci} we carried out a general study of the soft SUSY-breaking terms arising under the assumption
of Kahler moduli dominated SUSY-breaking in string theory. Under the additional assumption of a unified 
structure analogous to that  obtained in $SU(5)$ orientifolds or F-theory GUT's one obtains universal
soft parameters, similar to those in the CMSSM  
or slight generalizations.  Imposing correct radiative electroweak symmetry breaking (REWSB) \cite{ir2}
and viable neutralino dark matter we found that essentially only one single type of configuration survives. 
These are models in which the SM fields live  at the intersection of 7-branes, very much like in the
recent F-theory GUT constructions (see Refs.\,\cite{weigandRev,heckmanRev,Wijnholt:2010zz}
for reviews and references). In the latter, quarks and leptons live confined in complex 
matter curves embedded in the bulk 7-brane in which the SM gauge group lives (see Fig.\,\ref{ftheoryguts}).
Yukawa couplings arise at the intersection points of the different matter curves.
It must be emphasized that this kind of constructions form a large class, since several other string constructions are
their duals. Thus for example Type IIA orientifolds with the SM at intersecting D6-branes are their mirror and  F-theory
constructions are also directly related to M-theory compactifications in manifolds of $G_2$ holonomy, see
Ref.\,\cite{book} for a review of these connections.

  In the present paper we explore in further detail this string theory configuration beyond the results of Ref.\,\cite{aci}		
  and study  its phenomenological consequences,  including the Higgs masses and sparticle spectrum.
  We also study the LHC reach in testing these models. In doing this analysis we find a number of  interesting new  results:
  
  \begin{itemize}
  
  \item
  We have realized that our construction, put forward  a few years ago \cite{aci},  does contain the ingredients which
  favour a relatively heavy lightest CP-even Higgs mass. Indeed, in these constructions one has a very predictive set
  of boundary conditions with $M=\sqrt{2}m= -(2/3)A=-B$ so that one is essentially left with two free parameters, $M$ and $\mu$.
  In particular, this implies $A=-3/\sqrt{2}m\simeq -2m$, {\em one of the necessary conditions for a heavy lightest Higgs}
  (for not too heavy squark/gluino spectrum).

 \item 
The boundary conditions are consistent with radiative EW symmetry breaking  and a slight deformation (which might be induced by gauge fluxes and which leads to non-universal Higgs mass parameters) makes it also compatible with viable neutralino (mostly bino) dark matter. 
The correct neutralino relic abundance is obtained through coannihilation effects when the neutralino, $\chi_1^0$,  and the lightest stau, ${\tilde \tau}_1$, are almost degenerate in mass.
However, {\em this only happens for large values of $\tan\beta\simeq40$} (the second condition for a heavy Higgs), and  $ M\leq 1400\,{\rm GeV}$, where $M$ is the universal gaugino mass at the unification scale.
There is also  a lower bound $M\geq 570$~GeV in order not to violate the experimental bounds on the branching ratio of the rare process  $b\rightarrow s\gamma$.
Since this construction has essentially three free parameters, $M$, $\mu$ and a small flux parameter $\rho_H$, after imposing REWSB and correct dark matter the Higgs and sparticle spectrum are  very much constrained. In particular, having $570\,{\rm GeV}\leq  M \leq 1400\,{\rm GeV}$ forces the lightest Higgs mass to be in the range 
\begin{equation}
	119 \,{\rm GeV} \ \leq m_h \  \leq 125 \,{\rm GeV}  \ ,
\end{equation}
in agreement with the range favoured by 2011 LHC data.
In addition, the recent constraint on BR($B_s\to\mu^+\mu^-$) reported by the LHCb collaboration \cite{Aaij:2012ac}, further reduces this range, leaving only the region with $m_h\sim125$~GeV.

\item
Once the Higgs  mass is known, one can extract, at least in principle,  the complete SUSY and Higgs mass spectrum. In practice both the experimental error  of the Higgs and the top quark mass as well as inherent errors on the computation of the Higgs mass somewhat limit the accuracy of this computation.
Still, this range of Higgs masses points to a relatively heavy SUSY spectrum but, fortunately, testable at LHC.
A Higgs heavier than $119$~GeV would imply squarks and gluinos heavier than approximately
$1.2$~TeV, consistent with LHC limits obtained with 1 fb$^{-1}$. If the signal for a Higgs mass around $125$~GeV is real, one would expect first generation  squarks of order $2.8$~TeV and gluinos of order $3$~TeV.

\item In order to find the LHC reach for this model we have simulated background and signal events using Monte Carlo tools (PYTHIA and PGS). From the jets+missing energy signature,  we find that the LHC at 7(8) TeV will be able to test the model up to $M\leq 600(700)$ GeV, for an integrated luminosity around 20fb$^{-1}$, corresponding to squarks and gluino masses around 1.4~TeV. Likewise, the LHC at 14~TeV will be able to test the full parameter space with $M\leq 1.4$~TeV and an integrated luminosity around 25 fb$^{-1}$.

\item
In the region with $m_h\approx125$~GeV we find that the mass-difference between the lightest stau and neutralino masses is extremely small, $(m_{{\tilde \tau}_1}^2-m_{\chi_1^0}^2)\approx0.1$~GeV, thereby making the stau a very long-lived particle. Interestingly, the stau has the right properties to lead to Catalyzed BBN, alleviating the problems associated to the Lithium abundance in standard BBN.

\item 
CMSSM-like models are known to require a certain amount of fine-tuning and this is no exception.
A fine-tuning of order of a percent in the $M$, $\mu$ and $\rho_H$ parameters is expected  in order to obtain
both correct REWSB and viable neutralino dark matter. Concerning the origin of this fine-tuning,
the fact that small deviations from the parameters drive the theory into {\it catastrophic} regions with unbroken EW 
symmetry and/or above critical matter densities suggest a possible environmental
(anthropic) explanation. It has been argued 
 that the  {\it little hierarchy} of the MSSM  could be a result of an anthropic selection \cite{Giudice:2006sn}.  
 We argue that the requirement of viable neutralino dark matter could also add arguments in that direction. 
 One may argue that if the free parameters {\it scan} in a landscape, this would tend to favor the heaviest
 spectrum consistent with {\it both} REWSB and neutralino dark matter. This would in turn favor the largest
  Higgs mass  within this scheme, of order 125 GeV.

\end{itemize}

The paper is organised as follows. In the next chapter we give a brief overview of the 
soft terms which are induced by modulus dominance in Type IIB models with the SM fields at intersecting
7-branes. In chapter 3 we study the Higgs and sparticle spectrum consistent with both REWSB and 
viable dark matter in the context of this model. In chapter 4 we study the LHC reach for testing  the model and in
chapter 5 we discuss the possible environmental origin of the fine-tuning required in this class of models.
Conclusions are left for chapter 6.

\section{SUSY-breaking in string theory and modulus dominance in Type IIB orientifolds }

We present here a brief review of a few 
elements which are relevant for the construction of this class of models, see
 Refs.\,\cite{book} and \cite{aci} for further  details. Readers interested only in the phenomenological applications 
 of the model may jump safely to chapter 3.
We  assume that the SM gauge fields reside at a stack of 7-branes wrapping a   4-cycle  $S$ (of size controlled by
a Kahler modulus $t$) in
a  6-manifold whose overall volume is controlled by a large modulus $t_b\gg t$. In the F-theory context these
moduli $t,\,t_b$ would correspond to the size of $S$ and $B_3$.
As argued  in Ref.\,\cite{aci} we can model out this
structure with a Kahler potential of the form \cite{Balasubramanian:2005zx,Conlon:2005ki}
\beq
G \ =\ -2\log(t_b^{3/2} \ -\ t^{3/2}) \ +\ \log|W|^2\,,
\eeq
with $t=T+T^*$ being the relevant local modulus associated to the SM and $W$ the full superpotential \footnote{We ignore the dependence on
the dilaton and complex structure fields which are typically fixed in the presence of closed string fluxes.
There may also be additional Kahler moduli which will not modify the general arguments applicable to any
local brane configuration.}.
The SM matter fields $C_\alpha$ of the MSSM reside 
at  the intersection of 7-branes. Then the gauge kinetic function and the Kahler metrics of the
matter fields are given to leading order in $1/t_b$  by \cite{Conlon:2006tj,aci}
\beq
f\ =\ T \ ;\ K_\alpha \ =\ \frac {t^{1-\xi_\alpha}}{t_b} \,,
\eeq
where $\xi_\alpha$ is the {\it modular weight} of the corresponding particle.
Its value depends on the geometrical origin of the field with $\xi_\alpha=1/2$ for fields
localized on intersecting 7-branes.
Note that the SM gauge couplings are unified and determined by the real part of $f$.
Using this information and assuming that the auxiliary field of the local modulus has $F_T\not= 0$, 
using standard supergravity formulae (like e.g. those in Ref.\,\cite{bim}) 
it is easy to derive the 
simple set of  soft term boundary conditions \cite{fluxed,aci}
\beqa
m_\alpha^2\ & =& \ (1-\xi_\alpha)|M|^2 \ ,\ \alpha=Q,U,D,L,E,H_u,H_d \, , \\ \nonumber
A_U \ &\ =&  \ -M(3\ -\ \xi_{H_u}\ -\ \xi_Q \ -\ \xi_U)\,, \\ \nonumber
A_D \ &\ =&  \ -M(3\ -\ \xi_{H_d}\ -\ \xi_Q \ -\ \xi_D)\,, \\ \nonumber
A_L \ &\ =&  \ -M(3\ -\ \xi_{H_d}\ -\ \xi_L \ -\ \xi_E)\,, \\ \nonumber
B \ & = & \ -M(2\ -\ \xi_{H_u} \ -\ \xi_{H_d})\,,
\label{modularsoftterms}
\eeqa
where $M$ is the universal gaugino mass and the notation is standard. In the case under consideration
quarks, leptons and Higgs fields live at 7-brane intersections and hence $\xi_\alpha=1/2$ for all $\alpha$.
Then one gets the simple set of boundary conditions
\beq
M\ =\ \sqrt{2}m\ = \ -(2/3)A\ =\ -B \, .
\label{semboundary}
\eeq
Here we have assumed that there is an explicit $\mu$-term from some unspecified origin (possibly also fluxes),
so that the model would have in principle only two free parameters, $M$ and $\mu$  and therefore constitutes
a {\it slice of the CMSSM} boundary conditions.

In general, magnetic flux backgrounds may be present on the worldvolume of
the 7-branes in order to get a chiral spectrum. 
In the presence of
magnetic flux backgrounds in the 7-branes the kinetic functions and Kahler metrics may 
get small corrections which have the form in the dilute flux approximation \cite{Lustflux,aci}
\beq
f\ =\ T( 1\ + \ a \frac{S}{T} )\ \ ;\ K_\alpha \ =\ \frac {t^{1/2}}{t_b} (1\ +\  \frac{c_\alpha}{t^{1/2} } )\ ,
\eeq
where $a$ and $c_\alpha$ are constants and $S$ is the  the complex dilaton field.
These  corrections are suppressed in the large $t$ limit, corresponding to the physical weak coupling.
In this limit one may also neglect the correction to $f$ compared to that coming from $K_\alpha$.
One then finds corrected soft terms of the form
\beqa
m_{\tilde f}^2 \  & = & \ \frac{1}{2}|M|^2(1-\frac {3}{2}\rho_f)\,, \\  
m_{H}^2 \ & = & \   \frac{1}{2}|M|^2  (1- \frac{3}{2}\rho_H)\,,  \\
A \ & = & \ -\frac{1}{2}M(3\ -\ \rho_H\ -\ 2\rho_f)\,, \\ 
B\ & = & \ -M(1-\rho_H)  \, ,
\label{boundconditionsfinal}
\eeqa
where $\rho_\alpha=c_\alpha/t^{1/2}$.  Note that as an order of magnitude one numerically expects 
$\rho_H\simeq 1/t^{1/2}\simeq \alpha_{GUT}^{1/2}\simeq 0.2$. These expressions are further  simplified if one assumes that,
e.g.,  only the flux correction to the Higgs Kahler metric  is non negligible. This is for example what happens 
in F-theory $SU(5)$ GUTs, in which it is assumed that the hypercharge flux is only non-vanishing in the
Higgs matter curve.  In what follows we will only consider this case, although we have done an analogous
analysis with $\rho_f\not=0$ which yields completely analogous results (although requiring slightly larger $\rho_H$).

\section{Higgs and SUSY  spectrum  in the  Modulus Dominated CMSSM} 

\begin{figure}[h!]
\hspace*{-0.6cm}
\centering
\includegraphics[width=8.cm, angle=0]{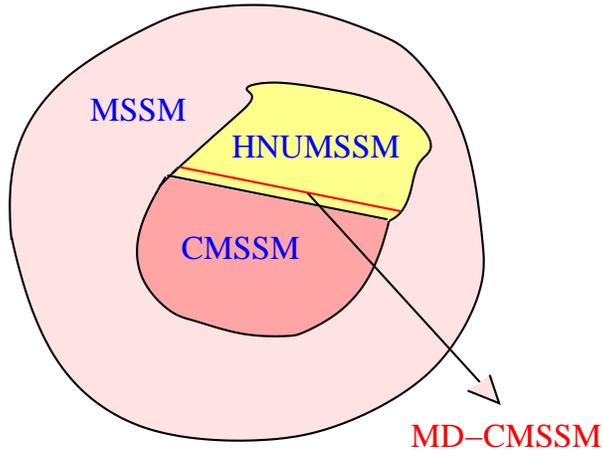}
\caption{Pictorial view of the modulus dominance constrained MSSM as a
slice of the Higgs non-universal HNUMSSM which is a slight deformation 
(due to the small flux parameter) of the CMSSM.}
\label{MDCMSSM}
\end{figure}

In the scheme under discussion we are thus left with soft terms at the string unification scale 
with the relations
\beqa
m_{\tilde f}^2 \  & = & \ \frac{1}{2}|M|^2\,, \\  
m_{H}^2 \ & = & \   \frac{1}{2}|M|^2  (1- \frac{3}{2}\rho_H)\,,  \\
A \ & = & \ -\frac{1}{2}M(3\ -\ \rho_H)\,, \\ 
B\ & = & \ -M(1-\rho_H)\,, 
\label{boundconditionsfinal2}
\eeqa
where $\rho_H$ parametrizes the effect of magnetic fluxes on the Higgs Kahler metrics, see Ref.\,\cite{aci}.
As we  said, this set of soft terms  constitutes a  deformation of a slice of the CMSSM with slightly non-universal
Higgs masses. We will call it Modulus Dominated CMSSM (MD-CMSSM,  Fig.\,\ref{MDCMSSM}).

Consistency of the scheme
requires this parameter to be small so that indeed the interpretation of $\rho_H$ as a small flux correction makes sense.
Note that we thus have essentially two free parameters, $M$ and $\mu$, with a third parameter $\rho_H$ restricted to be small.
We are going to impose two constraints: 1) consistent REWSB and 2) correct neutralino dark matter abundance. These two constraints
are very stringent and it is non-trivial that both conditions may be simultaneously satisfied in such a constrained system \cite{aci}.

\subsection{REWSB and dark matter constraints: a model with a single free parameter}

We have performed a detailed analysis of both REWSB and dark matter constraints based on the above boundary conditions. 
A similar study  was made  in Ref.\,\cite{aci} but here we carry out a more through analysis, covering the full parameter space
and studying in detail the Higgs and SUSY spectra. We also analyze the impact of the 2011 LHC data on our results and explore the eventual LHC reach in testing the model.

The minimization condition of the effective Higgs potential gives rise to the 
weak scale equation
\beq
  {\mu^2} \,= \,
  \frac{ -  {m_{H_u}^2} {\tan^2\beta} +  {m_{H_d}^2}}{ {\tan^2\beta}-1}
  - \frac{1}{2}  {M_Z^2 } \, ,
  \label{condicionrewsb}
  \eeq
 with 
\beq
\sin2\beta\ =\ \frac {2|B\mu|}{(m_{H_u}^2+m_{H_d}^2+2\mu ^2)} \  ,\quad  \tan\beta\equiv\nu_u/\nu_d\,. 
\label{relmubeta}
\eeq
 In principle, the usual procedure consists in fixing the value of $\tan\beta$ and then using Eq.\,(\ref{condicionrewsb}) to obtain the 
 modulus of $\mu$. The value of $B$ is then obtained from Eq.\,(\ref{relmubeta}). In our case the value of $B$  at the 
 unification scale is also predicted so that Eqs.\,(\ref{condicionrewsb}) and (\ref{relmubeta}) can be used to obtain both
 the values of $\tan\beta$ and $\mu$ in terms of a single parameter $M$ (plus the dependence on the small
 flux parameter $\rho_H$). Since it is not possible to derive an analytical solution for $\tan\beta$ from Eqs.\,(\ref{condicionrewsb}) and (\ref{relmubeta}),
 and given that we also need  the value of $\tan\beta$ to adjust the values of the Yukawa couplings at the unification scale, an
 iterative procedure has to be followed in which the RGE are solved numerically for a tentative value of $\tan\beta$, with the soft terms 
 given by Eqs.\,(\ref{boundconditionsfinal}) in terms of the two parameters $M$ and $\rho_H$.  The resulting $B$ at the weak
 scale is then compared to Eqs.\,(\ref{condicionrewsb}) and (\ref{relmubeta}), and the value of $\tan\beta$ is varied until agreement is reached.
 It is often not possible to find a solution with consistent REWSB and this excludes large areas of the ($M,\,\rho_H$) parameter space.
 
We have implemented this iterative process through a series of changes in the public code {\tt SPheno\,3.0}
 \cite{Porod:2003um,Porod:2011nf}. This code solves numerically the renormalization group equations of the MSSM and provides the SUSY spectrum at low energy. We use this code through a link in {\tt MicrOMEGAs 2.4} \cite{Belanger:2001fz,Belanger:2004yn,Belanger:2010gh}, which also calculates the theoretical predictions for low-energy observables such as the branching ratios of rare decays ($b\to s\gamma$, $B_s\to\mu^+\mu^-$) and the muon anomalous magnetic moment.
 The results are sensitive to the value of the top quark mass, particularly for the Higgs mass, see below. In the computation we use the central value in $m_t=173.2\pm 0.9$~GeV \cite{Lancaster:2011wr}.

In addition to correct REWSB we also impose the presence of viable neutralino dark matter, assuming R-parity conservation. The relic density of the neutralino is calculated numerically using the MSSM module of the code {\tt MicrOMEGAs 2.4} 
and we check for compatibility with the data obtained from the WMAP satellite, which constrain the amount of cold dark
 matter to be 
 $0.1008 \leq \Omega h^2\leq 0.1232$ at the $2\,\sigma$ confidence level \cite{Komatsu:2010fb}.

\begin{figure}[t!]
\hspace*{-0.6cm}
\includegraphics[width=8.5cm, angle=0]{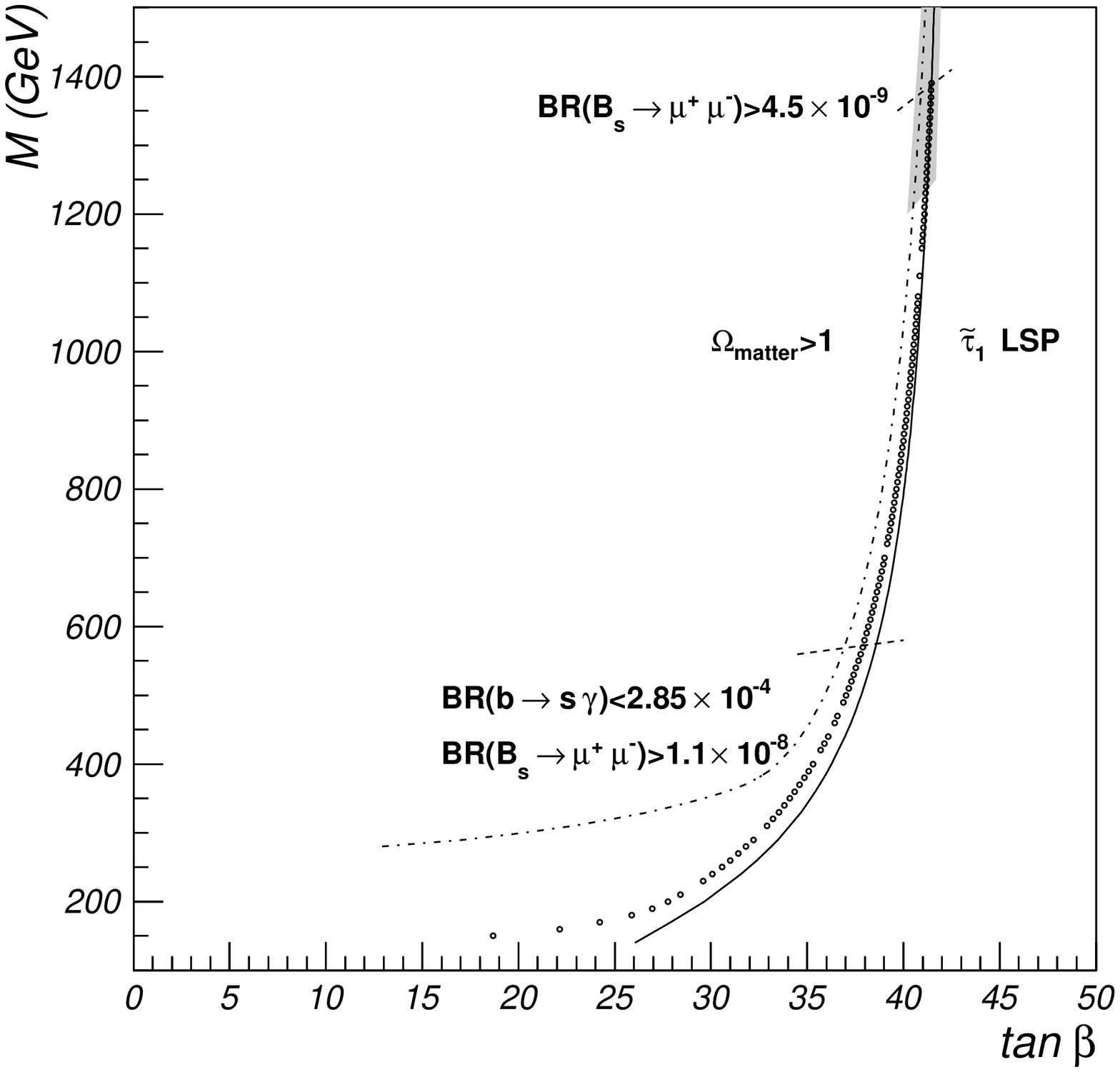}
\includegraphics[width=8.5cm, angle=0]{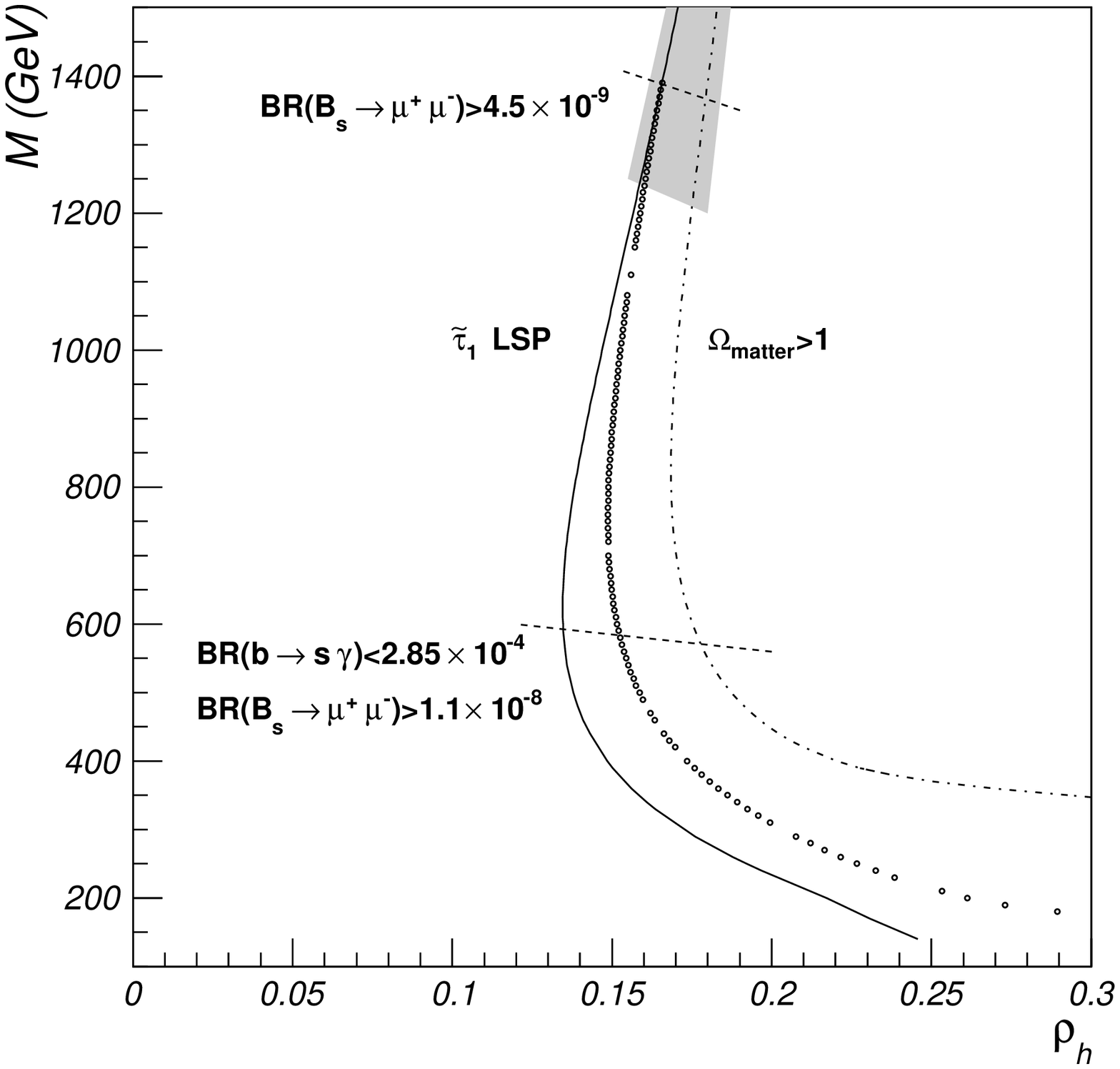} 
\caption{Left) Trajectory in the $(M,\tan\beta)$ plane for which the REWSB conditions are fulfilled and the correct amount of 
dark matter is obtained. Right) Corresponding values of the flux, $\rho_H$. In both cases, dots correspond to points fulfilling the central value in WMAP result for the neutralino relic density. The dot-dashed line denotes points along which the matter density is critical, $\Omega_{matter}=1$, whereas the solid line indicates the points for which the stau becomes the LSP. The points below the dashed line are excluded by the lower bound on BR$(b\to s\gamma)$ and the upper bound on BR$(B_s\to\mu^+\mu^-)$ from Ref.\,\cite{cmslhcb} and the recent LHCb result \cite{Aaij:2012ac}. The gray area indicates the points compatible with the latter constraint when the $2\sigma$ error associated to the SM prediction is included.}
\label{Mtanbeta}
\end{figure}

Imposing both conditions we are left with a model with a single free parameter or, equivalently, lines in the  
$(M,\,\tan\beta)$ and  $(M,\,\rho_H)$ planes.   In Fig.\,\ref{Mtanbeta} we show the trajectories consistent with both REWSB and viable
 neutralino dark matter. The left hand-side of Fig.\,\ref{Mtanbeta} shows how the viable values for $\tan\beta$ are confined to a large value region,
 $\tan\beta \simeq 36-41$. The maximum values for $M$ and $\tan\beta$ occur for $M\simeq 1.4$~TeV, $\tan\beta\simeq 41$. 
 The existence of these maximal values are due to the dark matter condition. Indeed, as we will see momentarily,  the LSP in
 this scheme is mostly pure Bino and generically its abundance exceeds the WMAP constraints. However along the line in the figure 
 the lightest neutralino $\chi_1^0$ is almost degenerate in mass with the lightest stau ${\tilde \tau}_1$  (see Fig.\,\ref{plotHiggs1})  and a coannihilation effect takes place
 in the early universe reducing very effectively its abundance. Above the point $M\simeq 1.4$ TeV, 
 coannihilation is not sufficiently efficient in depleting neutralino abundance and 
  $\chi_1^0$  ceases to be a viable dark matter candidate. Thus viable dark matter 
 gives rise to a very strong constraint on the $M$ value, $|M|\leq 1.4$~TeV, which in turn implies an upper bound on the SUSY and Higgs spectrum, see below. 
Notice that for small values of the gaugino mass the predicted $\tan\beta$ can also be smaller. In principle one could get to values of $\tan\beta$ as low as $~10$ while still fulfilling REWSB and the neutralino relic density with $M\gsim150$~GeV. However, the resulting SUSY spectrum is extremely light and already well below the current experimental bounds. First, demanding $m_h>115.5$~GeV leads to a lower bound on the common scale $M\gsim340$~GeV with $\tan\beta\gsim34$. 
Similarly, current LHC lower bounds on the masses of gluinos and second and third generation squarks imply $M\gsim400$~GeV and $\tan\beta\gsim35$. 
There is a  more stringent lower bound coming from the BR($b\to s\gamma$) constraint, which implies $M\gsim570$~GeV and $\tan\beta\gsim38$. 

\begin{figure}[t!]
\centering
\includegraphics[width=8.5cm, angle=0]{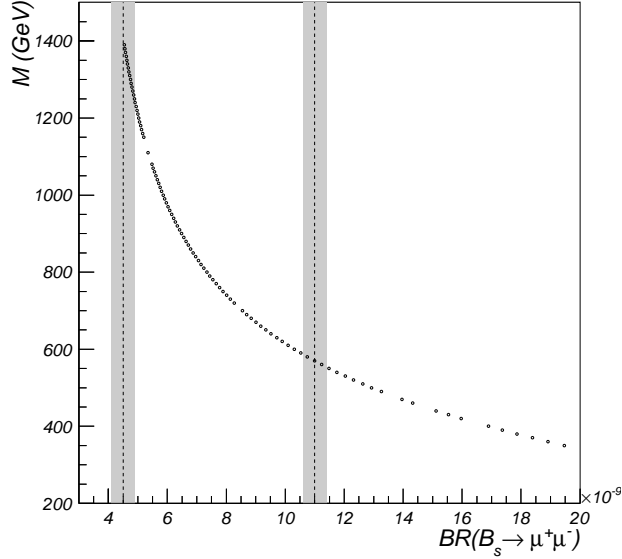}
\caption{Universal gaugino mass versus the theoretical prediction for BR$(B_s\to\mu^+\mu^-)$. The dashed lines denote the experimental upper bound on this observable from Ref.\,\cite{cmslhcb} and the recent LHCb result \cite{Aaij:2012ac}. The $2,\sigma$ theoretical error on the SM prediction is indicated by means of a shaded region in both cases.}
\label{fig:bmumu}
\end{figure} 

Finally, the experimental upper limit on BR($B_s\to\mu^+\mu^-$) has a profound impact on the allowed parameter space.
A combination of CMS \cite{Chatrchyan:2011kr} and LHCb \cite{lhcb} data recently set a bound as low as BR$(B_s\to\mu^+\mu^-)<1.1\times10^{-8}$ \cite{cmslhcb}. This would lead to $M\gsim560$~GeV, thus having a similar effect as the other constraints mentioned above.
However, as this work was released, the experimental bound was significantly improved by the LHCb collaboration \cite{Aaij:2012ac}, leading to the unprecedented constraint BR$(B_s\to\mu^+\mu^-)<4.5\times10^{-9}$. This is in fact very close to the SM prediction BR$(B_s\to\mu^+\mu^-)=(3.2\pm0.2)\times10^{-9}$ \cite{Buras:2010mh,Buras:2010wr} and thus
has important implications in our parameter space. Given that our model entails large values of $\tan\beta$ and a significant mixing in the stop mass matrix, the resulting BR$(B_s\to\mu^+\mu^-)$ is relatively large. 
Fig.\,\ref{fig:bmumu} represents the theoretical predictions for this observable as a function of the corresponding universal gaugino mass, showing that BR$(B_s\to\mu^+\mu^-)\gsim4.4\times10^{-9}$. We display in the plot the experimental bound from Ref.\,\cite{cmslhcb} and Ref.\,\cite{Aaij:2012ac}, explicitly showing the effect of the improved measurement. For each case, we take into account the $2\sigma$ theoretical uncertainty of the SM contribution.
It is in fact expected that this upper bound improves in the near future with new data from CMS and LHCb. This has the potential to disfavour our construction if no deviation from the SM value is observed.Ê
\footnote{It should be pointed out in this respect that the inclusion of non-vanishing flux correction $\rho_f$  for sfermions in Eq.\,(\ref{boundconditionsfinal})
can slightly alter the allowed regions in the parameter space, shifting the viable points towards smaller values of $\tan\beta$, thereby decreasing the SUSY contribution to BR$(B_s\to\mu^+\mu^-)$.}

On the right hand-side of Fig.\,\ref{Mtanbeta} we display the line in the $(M,\,\rho_H)$ plane that is consistent with REWSB and viable neutralino dark matter.
Interestingly enough, after applying experimental constraints, the value of $\rho_H$ is indeed small, of order $0.15-0.17$ and is very weakly dependent on $M$.
This is consistent with the interpretation of $\rho_H$ as a small correction arising from gauge fluxes, as discussed in the previous chapter.
Indeed the values for $\rho_H$  obtained are of the expected order of magnitude, $\rho_H\propto \alpha_{GUT}^{1/2}\simeq 0.2$.
 
The viable points of the parameter space lie along a narrow area of the parameter space. In fact, small deviations in any of the parameters, $M$, $\tan\beta$ or $\rho_h$ have catastrophic consequences, since either the relic density becomes too large (it very rapidly overcloses the Universe) or the stau becomes the LSP. We illustrate this in Fig.\,\ref{Mtanbeta}, where the dashed and solid lines represent the points for which $\Omega_{matter}=1$ and $m_{{\tilde \tau}_1}=m_{\chi_1^0}$, respectively.
The line with critical density extends to $M\approx2.5$~TeV, but the region fulfilling WMAP $2\,\sigma$ region stops at $M=1.4$~TeV. 
Interestingly, the flux $\rho_h$ cannot vanish (since the stau becomes the LSP), this is, even though small, a deviation from the CMSSM is necessary. Also, it cannot be too large or we would have an excessive amount of dark matter.

\begin{figure}[t!]
\hspace*{-0.6cm}
\centering
\includegraphics[width=11.cm, angle=0]{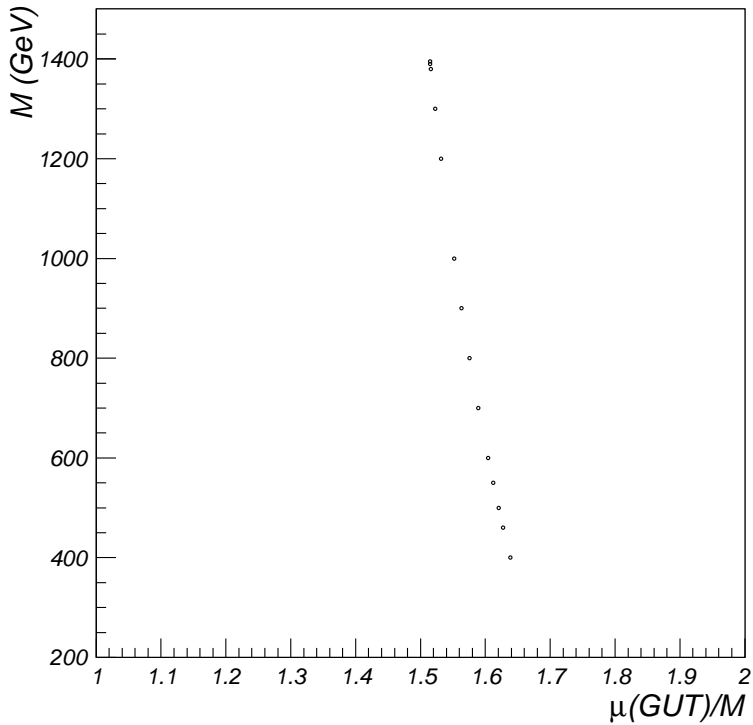}
\caption{Universal gaugino mass as a function of the resulting Higgsino mass parameter, $\mu$ at the unification  scale for those points where both REWSB and viable neutralino dark matter are obtained.}
\label{plotmu}
\end{figure} 
 
As we explained in  the beginning of this chapter, the $\mu$ parameter is computed at the electroweak scale from Eq.\,(\ref{condicionrewsb}). 
Using {\tt SPheno\,3.0} we have also computed its value at the unification scale (the effect of the RGEs is not large for this parameter) so that we can compare it with the soft parameters. This might give us an idea of what the possible origin of the $\mu$-term could be
\footnote{In particular, as  noted in Ref.\,\cite{aci},  the Giudice-Masiero  \cite{gm} mechanism would predict in the present model
$\mu=-M/2$ and $B=-3M/2$, which do not lead to consistent REWSB.}.
The results are displayed in Fig.\,\ref{plotmu}, where the ratio $\mu({\rm GUT})/M$ that corresponds for each value of the gaugino mass is plotted. As we can observe, the predicted value for that ratio is approximately constant and satisfies 
$\mu\sim(1.5-1.6)\,M$. At the point of maximal $M$ one has approximately $|\mu|=|A|=3/2|M|$.
This could perhaps point towards a higher degree of interdependence among  soft terms, see the discussion in chapter 5.

\subsection{The Higgs mass}

The lightest neutral Higgs, $h$, in the MSSM receives important one-loop corrections to its mass from the top-stop loops. 
The one-loop corrected Higgs mass has an approximate expression of the form
\cite{higgsloops}
\beq
m_h^2\ \simeq \ M_Z^2\cos^22\beta \ +\ 
\frac {3m_t^4}{16\pi^2v^2} \left( \log\frac{m_{\tilde t}^2}{m_t^2} \ +\ 
\frac{X_t^2}{m_{\tilde t}^2}    \left(1-\frac{X_t^2}{12m_{\tilde t}^2}\right)   \right) \ ,
\eeq
where $v^2=v_1^2+v_2^2$, $m_{\tilde t}=(m_{{\tilde t}_L}m_{{\tilde t}_R})^{1/2}$, and $X_t=A_t-\mu \cot\beta$,
all evaluated at the weak scale.
The largest values for the Higgs mass are obtained then for large $\tan\beta$ and large stop masses. 
In particular, the quantity in brackets is maximized for $|X_t|\simeq\, \sqrt{6}m_{\tilde t}$. 
Interestingly enough this maximal value typically correspond to large values for the trilinear soft term 
$A/m\simeq \pm 2$ (see e.g. Ref.\,\cite{Baer:2011ab,Arbey:2011ab}). In our scheme we have  $A/m=-3/\sqrt{2}+\rho_H/\sqrt{2}\simeq -2$
and large values of $\tan\beta=36-41$, so that relatively  {\it large values of the Higgs mass are an automatic
prediction of our scheme}.

\begin{figure}[t!]
\hspace*{-0.6cm}
\centering
\includegraphics[width=11.cm, angle=0]{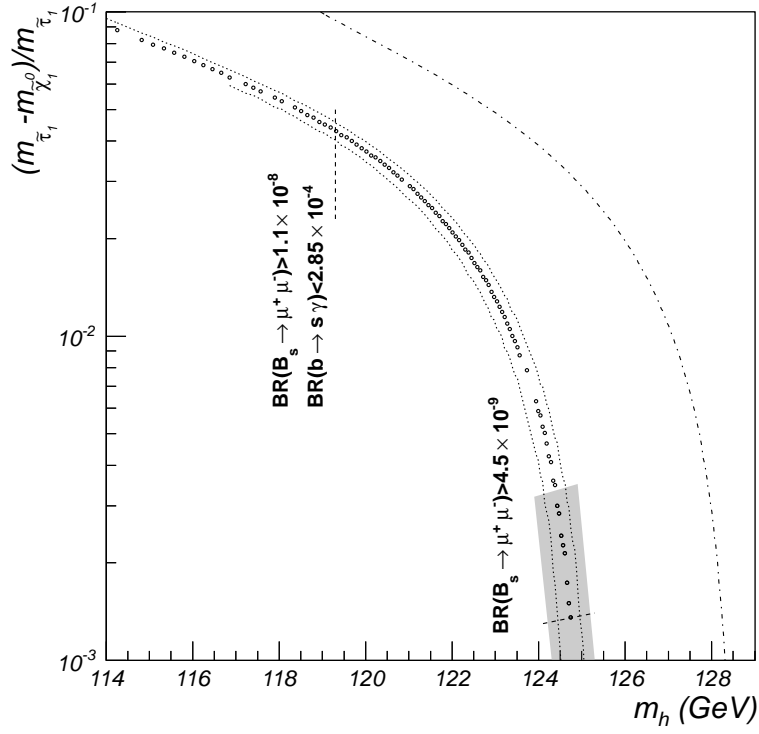}
\caption{The normalized mass difference 
$(m_{{\tilde \tau}_1}-m_{\chi_1^0})/m_{{\tilde \tau}_1}$ as a function of the lightest Higgs mass $m_h$.  Dots correspond to points fulfilling the central value in the result from WMAP for the neutralino relic density and dotted lines denote the upper and lower limits after including the $2\sigma$ uncertainty. 
The dot-dashed line represents points with a critical matter density $\Omega_{matter}=1$. The vertical line corresponds to the 2$\sigma$ limit on BR($b\rightarrow s\gamma$) and the upper bound on BR$(B_s\to\mu^+\mu^-)$ from Ref.\,\cite{cmslhcb} and the recent LHCb result \cite{Aaij:2012ac}. The gray area indicates the points compatible with the latter constraint when the $2\sigma$ error associated to the SM prediction is included.} 
\label{plotHiggs1}
\end{figure}

The 2011 run at LHC has restricted the most likely range for a SM Higgs to the range $115.5-131$~GeV (ATLAS) and 
$114.5-127$~GeV (CMS). Furthermore there is an excess of events in the  $\gamma \gamma$, $ZZ^*\rightarrow 4l$ and 
$WW^*\rightarrow 2l$ channels suggesting the presence of a Higgs boson at a mass around $125$~GeV. 
Although more data are needed to confirm this excess, it is interesting to see whether a Higgs boson in that 
range appears in this construction.  
As we said, our scheme has essentially one free parameter and the 
allowed values for the Higgs mass turn out to be very restricted. We have computed the mass of the Higgs particles
to two-loop order 
using the code {\tt SPheno} linked through the {\tt micrOMEGAs} program
\footnote{We have compared our results with those obtained with {\tt FeynHiggs\,2.8.6} \cite{Heinemeyer:1998yj,Hahn:2005cu}, finding good agreement, within approximately 1~GeV.}.
To show the allowed values for the lightest Higgs mass we display in Fig.\,\ref{plotHiggs1} the ratio
$(m_{{\tilde \tau}_1}-m_{\chi_1^0})/m_{{\tilde \tau}_1}$  versus the value of
the lightest Higgs mass $m_h$. This mass difference is very relevant for the coannihilation effect which is required 
in this scheme to get viable neutralino dark matter.
We also illustrate the variation resulting from the $2\sigma$ uncertainty in the WMAP result.

One observes that there is a maximum value of the Higgs mass of order 125 GeV.  For higher values the
neutralino ceases to be viable as a dark matter candidate.  This limit corresponds to the
maximum allowed values $M\simeq 1.4$ TeV and $\tan\beta\simeq 41$ that we discussed above and hence
to a quite massive SUSY spectra, see below. There is also a lower limit coming from the 
lower bound on the constraint BR($b\rightarrow s\gamma)<2.85\times 10^{-4}$ 
which leads to
\beq
119\ {\rm GeV} \leq  \ m_h\ \leq 125\ {\rm GeV}  \ .
\eeq
In the MSSM the bound on BR$(B_s\to\mu^+\mu^-)$ also has an impact on the predicted Higgs mass \cite{Cao:2011sn}.
In our case, if the current LHCb constraint is taken at face value and the SM uncertainty is included in our theoretical predictions, the resulting range for the Higgs mass is reduced to
\beq
124.4\ {\rm GeV} \leq  \ m_h\ \leq 125\ {\rm GeV}  \ .
\eeq

We have to remark at this point that these values are sensitive to the value taken for the top quark mass and the corresponding error.
As we said we have taken the central value in $m_t=173.2\pm 0.9$  \cite{Lancaster:2011wr}.
The value of the Higgs mass is very dependent on the top mass.
As a rule of thumb, one can consider that an increase of $1$~GeV in the top mass leads to an increase of approximately $ 1$~GeV in $m_{h}$ \cite{Heinemeyer:1999zf}.
Note  also that the computation of the Higgs mass includes additional intrinsic errors 
of order 1 GeV, see e.g. Refs.\,\cite{Degrassi:2002fi,Allanach:2004rh}. In any event, it is remarkable that the allowed region in our model
is well within the range allowed by the 2011 LHC data. In particular, generic points in the CMSSM space tend to have a  lighter 
Higgs mass tipically of order 115 GeV or lower. Our particular choice of soft terms plus the constraint of viable neutralino
dark matter force our Higgs mass to be relatively high.

\begin{figure}[t!]
\hspace*{-0.6cm}
\centering
\includegraphics[width=8.5cm, angle=0]{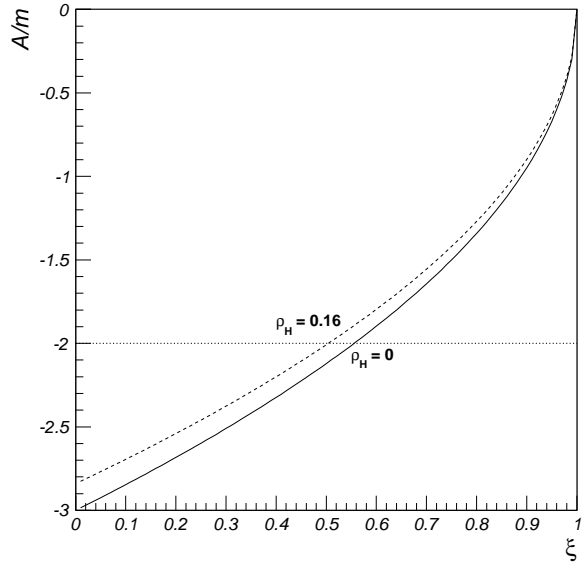}
\caption{Ratio $A/m_f$ at the GUT scale as a function of the modular weight $\xi$ for the case without fluxes (solid line) and when a small flux ($\rho_H=0.16$) is introduced.}
\label{aoverm}
\end{figure}

 It should be pointed out that the regions of the parameter space with larger values of the Higgs mass correspond to a heavy spectrum and therefore predict a small supersymmetric contribution to the muon anomalous magnetic moment, $a_\mu^{\rm SUSY}$. In particular, the points with $m_h>124$~GeV predict $a_\mu^{\rm SUSY}\approx3\times10^{-10}$.
These values show some tension with the observed discrepancy between the experimental value \cite{g-2} and the Standard Model predictions using $e^+e^-$ data, which imply $10.1\times10^{-10}<a_\mu^{SUSY}<42.1\times10^{-10}$ at the $2\,\sigma$ confidence level \cite{Hagiwara:2011af} where theoretical and expreimental errors are combined in quadrature (see also Refs.\,\cite{Jegerlehner:2009ry,Davier:2010nc}, which provide similar results). However, if tau data is used this discrepancy is smaller $2.9\times10^{-10}<a_\mu^{SUSY}<36.1\times10^{-10}$ \cite{Davier:2010nc}.

As we said, in the context of the CMSSM obtaining a large Higgs mass and not too heavy 
SUSY spectrum requires having $A\simeq -2m$. This may be
considered as a hint of a scheme with all SM localized in intersecting branes and is in fact independent of what the
possible origin of the $\mu$ term is. Indeed, for general (but universal) 
modular weights $\xi$ one has the relation
\beq
A\ = \ -3(1-\xi)^{1/2}\ m \ .
\label{estupenda}
\eeq
For $A/m\simeq -2$  one has $\xi\simeq 0.5$, indicating that indeed large Higgs masses favour 
all SM particles with $\xi\simeq 1/2$ modular weights, which correspond to intersecting branes,
as in our scheme. This is illustrated in Fig.\,\ref{aoverm}.

\subsection{The SUSY spectrum}

Again, our particular choice of soft terms significantly constrains the spectrum of SUSY particles.
Given that there is only one free parameter, fixing any value for a SUSY particle or Higgs field automatically 
fixes the rest of the spectrum. We give in Table\,\ref{espectromasas} the values of some masses and parameters as we vary 
the universal gaugino mass, $M$.  Let us remember that $\tan\beta$ is not an input, as it is fixed by the boundary conditions on $B$.
\begin{table}[t!] \footnotesize
\renewcommand{\arraystretch}{1.50}
\begin{center}
\begin{tabular}{|c||c|c|c|c|c|c|c|c|c|}
\hline 
M  &  $\tan\beta$ 
 &  ${\tilde g}$    &  ${\tilde Q}_L$
 &   ${\tilde Q}_R$     &    ${\tilde t}_1$  &   $\chi_2^0,\chi_1^+;L_R$   &    $\chi_1^0,{\tilde \tau}_1$        &    $M_A$ &   $m_h$   \\
\hline\hline
 $400$ & $35.3$ &
944   &  $900$  &   $870$    &   $605$    &    314;323  &  $164,175$ &     549 & 116.8 
 \\
 \hline
  $500$ & $37$ &
1160   &  $1107$  &   $1067$    &   $754$    &    397;402  &  $208,219$ & 660 &  118.5
 \\
 \hline
 $600$ & $38.2$ &
1372   &  $1310$  &   $1262$    &   $901$    &    481;482  &  $252,262$ & 769 &  119.7
 \\
 \hline
 $700$ & $39$ &
1583   &  $1511$  &   $1455$    &   $1046$    &    565;561  &  $296,305$ & 875 &  120.8
 \\
 \hline
 $800$ & $39.6$ &
1791   &  $1710$  &   $1644$    &   $1189$    &    649;641  &  $341,349$ & 981 &  121.7
 \\
 \hline
 $900$ & $40.1$ &
1998   &  $1907$  &   $1834$    &   $1330$    &    732;720  &  $386,393$ &  1084 &  122.4
 \\
  \hline
 $1000$ & $40.5$ &
2203   &  $2103$  &   $2020$    &   $1470$    &    816;800  &  $431,436$ &  1187 &  123.1
 \\
  \hline
 $1100$ & $40.8$ &
2424  &  $2314$  &   $2220$    &   $1620$    &    907;886  &  $480,483$ & 1298 & 123.7
 \\
   \hline
 $1200$ & $41.1$ &
2610  &  $2491$  &   $2390$    &   $1746$    &    984;859  &  $521,524$ & 1391 &  124.2
 \\
  \hline
 $1300$ & $41.3$ &
2812  &  $2683$  &   $2575$    &   $1883$    &    1069;1039  &  $567,568$ & 1492 & 124.7
 \\
   \hline
 $1400$ & $41.5$ &
3013 &  $2876$  &   $2760$    &   $2018$    &    1153;1119  &  $612,612$ & 1592 &  125.1
 \\
\hline
\end{tabular}
\end{center} \caption{\small  Sparticle and Higgs masses in GeV and resulting value of $\tan\beta$ as a function of $M$.
Note that there is a maximum value for $M=1.4$~TeV where $\chi_1^0$ becomes degenerate with the lightest  stau,
as the third column from the right shows. At that point the maximum value for the lightest Higgs mass $\simeq 125$~GeV
is obtained.
}
\label{espectromasas}
\end{table}

\begin{figure}[h!]
\hspace*{-0.6cm}
\centering
\includegraphics[width=10.cm, angle=0]{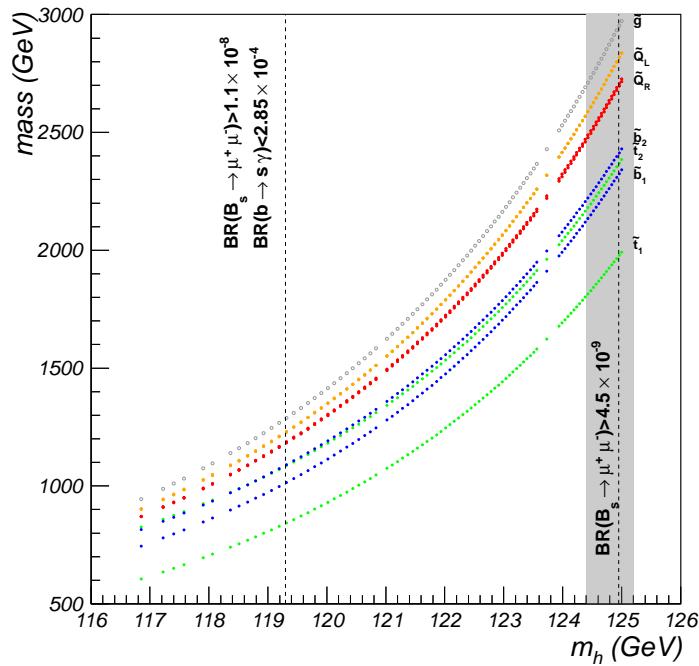} 
\caption{Squark and gluino masses as a function of the Higgs mass. 
The region to the left  of the vertical dashed indicates the constraint $BR(b\to s\gamma)<2.85\times 10^{-4}$ and the upper bound on BR$(B_s\to\mu^+\mu^-)$ from Ref.\,\cite{cmslhcb} and the recent LHCb result \cite{Aaij:2012ac}. The gray area indicates the points compatible with the latter constraint when the $2\sigma$ error associated to the SM prediction is included.}
\label{higgsversusquarks}
\end{figure}

One interesting way of presenting the structure of the SUSY spectrum is in terms of the lightest Higgs mass.
In Fig.\,\ref{higgsversusquarks} we show the masses  of the gluino and the squarks as a function of $m_h$.
The region to the left  of the vertical dashed line is excluded since it leads to  $BR(b\to s\gamma)<2.85\times 10^{-4}$.
Note that this implies that squarks of the first two generations and gluinos in our scheme must be heavier than $\simeq 1.2$~TeV.
This is consistent with LHC  limits obtained with 1\,fb$^{-1}$. For the third generation of squarks, the lightest stop 
has a mass of at least 800 GeV and the heaviest one, along with the sbottoms are heavier than 1 TeV.

If the signal for a Higgs at 125~GeV is real, one expects a quite heavy  spectrum with gluinos of order 3~TeV and squarks of the first two
generations of order 2.8~TeV. The lightest stop would be around 2~TeV and the rest of the squarks at around 2.3~TeV. 
Note however that these values depend strongly on the Higgs mass so that e.g. a Higgs around 124~GeV 
would rather correspond to squarks and gluinos around 2.2~TeV. Given the intrinsic error in the computation of
the Higgs mass, 
this only give us a rough idea of the expected masses for colored particles.
We discuss the testability of such  heavy colored spectra  in the next chapter.

\begin{figure}[h!]
\hspace*{-0.6cm}
\centering
\includegraphics[width=10.cm, angle=0]{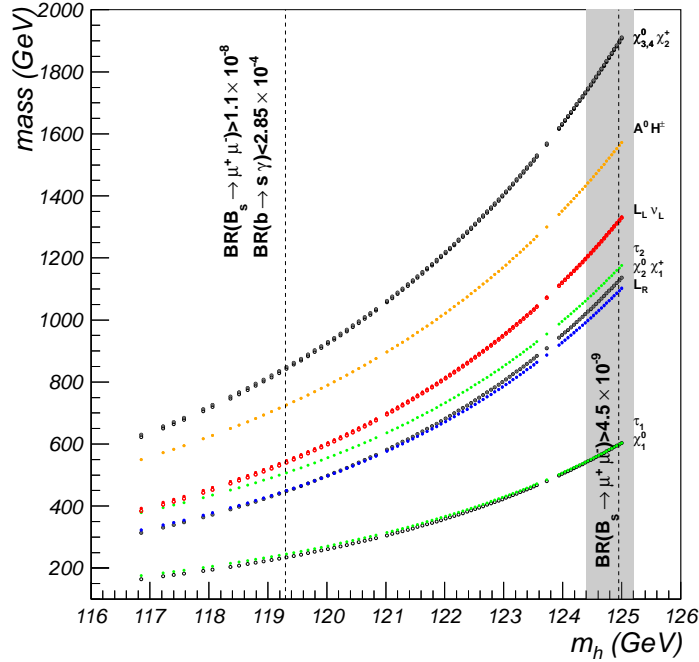}
\caption{Supersymmetric spectrum as a function of the Higgs mass of the slepton sector, together with the masses of the heavy Higgses and the gauginos.  The region to the left  of the vertical dashed indicates the constraint $BR(b\to s\gamma)<2.85\times 10^{-4}$ and the upper bound on BR$(B_s\to\mu^+\mu^-)$ from Ref.\,\cite{cmslhcb} and the recent LHCb result \cite{Aaij:2012ac}. The gray area indicates the points compatible with the latter constraint when the $2\sigma$ error associated to the SM prediction is included.}
\label{higgsversusleptons}
\end{figure}

In Fig.\,\ref{higgsversusleptons}  we show the spectrum of un-colored particles as a function of the lightest Higgs mass, including neutralinos, charginos, sleptons and 
the rest of the Higgs fields.  The region to the left  of the vertical dashed line is again excluded since  BR$(b\to s\gamma)<2.85\times 10^{-4}$.
The fact that $119$~GeV $\leq m_h\leq 125$~GeV strongly restricts the spectrum. 

The hierarchy in the sparticle mass pattern is a quick way of classifying a supersymmetric model and understanding the kind of signals it may give rise to in LHC. Several structures have been identified (see Ref.\,\cite{Feldman:2008hs} and references therein) that can originate from the CMSSM or non-universal supergravity scenarios. 
In our case, the model is very close to the CMSSM in the coannihilation region but further constrained. As a consequence, the resulting hierarchy in the supersymmetric spectrum is a very specific one. More specifically, the five lightest supersymmetric particles display the following structure:
\begin{eqnarray} 
\tilde{\chi}_1^0<\tilde{\tau}_1< \tilde{\chi}_2^0\approx\tilde{\chi}_1^\pm <\tilde l_R&{\rm for} & m_h<120\,{\rm GeV}
\,,\nonumber\\
\tilde{\chi}_1^0<\tilde{\tau}_1<\tilde l_R< \tilde{\chi}_2^0\approx\tilde{\chi}_1^\pm &{\rm for} & m_h>120\,{\rm GeV}
\,.\nonumber
\end{eqnarray}
These scenarios are analogous to mSP6 and mSP7, respectively, in Ref.\,\cite{Feldman:2008hs}.
The change of pattern is difficult to appreciate in Fig.\,\ref{higgsversusleptons}, since the mass difference between $\tilde l_R$ and the second-lightest neutralino is small (of order 10~GeV). Also, the mass difference between the second lightest neutralino and the lightest chargino is merely a fraction of a GeV. 

The almost identical values of the masses of $\chi_2^0$ and $\chi_1^\pm$ is expected since both fields are mostly Winos. 
On the other hand the degeneracy with the ${\tilde l}_R$ fields is a peculiarity of the structure of soft terms in this model. Indeed the 
weak scale masses for these fields have the structure
\beqa
M_{\chi_2^0}^2&  \simeq & \left(\frac{\alpha_2(M_Z)}{\alpha(M_s)}\right)M^2\simeq 0.64\, M^2\,, \\ \nonumber
m_{{\tilde l}_R}^2& \simeq & m^2\ +\ 0.15\,M^2\simeq 0.65\, M^2\,,
\eeqa
where in the second equation the boundary condition $m=M/\sqrt{2}$, characteristic of the present model,  has been used. 
From Fig.\,\ref{higgsversusleptons} we see that the lightest charged sparticle is a  stau, with a mass in between
200 and 550~GeV. The lightest slectrons and charginos are in the region 400 to 1000~GeV.  The remaining Higgs fields will
be heavy, in the 700 to 1600~GeV range. Thus there is a good chance to produce weakly interacting charged sparticles 
in a linear collider.

\begin{figure}[t!]
\hspace*{-0.6cm}
\centering
\includegraphics[width=11.cm, angle=0]{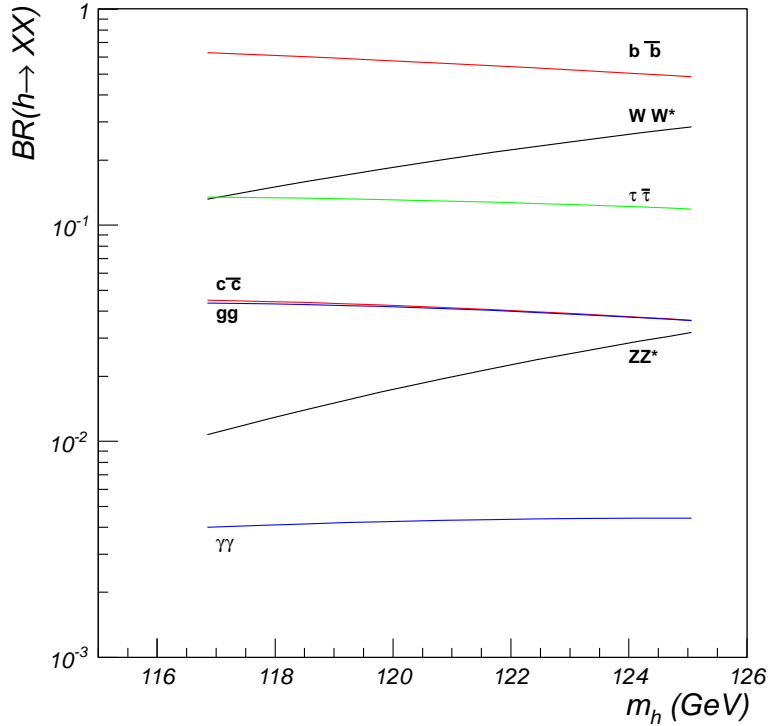}
\caption{Branching fractions for the decay of the lightest CP-even Higgs as a function of its mass.}
\label{plotHiggsbr}
\end{figure} 

For completeness we also show in Fig.\,\ref{plotHiggsbr} the branching ratios of the different decay modes of the lightest CP-even Higgs, computed using code {\tt SPheno\,3.0}, as a function of its mass in this construction. The composition of the lightest Higgs is very similar to that in the CMSSM and therefore these results are quite standard. The leading decay mode is $b\bar b$ although the contribution from $WW$ is almost comparable for large Higgs masses.

\begin{figure}[t!]
\hspace*{-0.6cm}
\centering
\includegraphics[width=11.cm, angle=0]{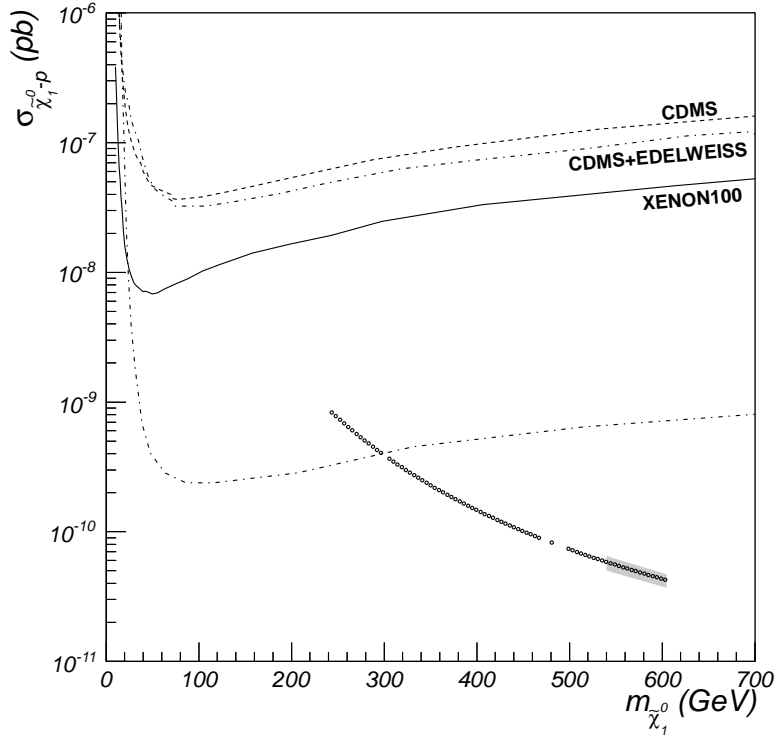}
\caption{Spin-independent part of the neutralino-proton cross section as a function of the neutralino mass for points reproducing the WMAP relic abundance and in agreement with all the experimental constraints. The current sensitivities of the CDMS \cite{Ahmed:2009zw}, CDMS combined with EDELWEISS \cite{Ahmed:2011gh} and XENON100 \cite{Aprile:2011hi} experiments are displayed by means of dashed, dot-dashed and solid lines, respectively. The dotted line represents the expected reach of a 1 ton experiment. The gray area indicates the points compatible with the LHCb constraint when the $2\sigma$ error associated to the SM prediction is included.}
\label{sigsi}
\end{figure}

Let us finally address the direct detectability of dark matter neutralinos in this construction. We show in Fig.\,\ref{sigsi} the theoretical predictions for the spin-independent contribution to its elastic scattering cross section off protons, $\crosssec$, as a function of the neutralino mass, together with current experimental sensitivities from the CDMS (showing also the combination of its data with those from EDELWEISS) and XENON detectors.  
After imposing all the experimental constraints, this scenario predicts $10^{-9}$~pb$\gsim\crosssec\gsim5\times10^{-11}$~pb. This is far from the reach of current experiments. 
Next generation experiments with targets of order 1 ton would be able to probe only 
a portion of the parameter space, corresponding to neutralino masses lighter than 300~GeV (and therefore to Higgses as heavy as approximately 121~GeV). This was to be expected, as these results are typical of the CMSSM in the coannihilation region. 

\section{Detectability at the LHC}

\subsection{Jets and missing transverse energy}

Having already described the SUSY spectrum, let us now address the detectability of this construction at the LHC. In the light of the current and predicted status of the collider, we will consider three possible configurations, with energies of $\sqrt{s}=7,\, 8$ and 14~TeV. 
Our goal is to determine the potential reach of the LHC as a function of the luminosity. In order to do so we have performed a Monte Carlo simulation for the different points in the viable parameter space.

As we commented in the previous chapter, the SUSY spectrum is calculated for each point using a modified version of {\tt SPheno\, 3.0}. The output, written in Les Houches Accord format, is directly linked to a Monte Carlo event generator. We have used {\tt PYTHIA 6.400} \cite{Sjostrand:2006za} to this aim, linked with {\tt PGS 4} \cite{PGS}, which simulates the response of the detector and uses {\tt TAUOLA} \cite{Jadach:1993hs} for the calculation of tau branching fractions.

We include the main sources for SM background, taking into account the production of $t\bar t$ and $WW/ZZ/WZ$ pairs, as well as $W/Z$+jets. The latter give the main contribution \cite{cmshadro,Aad:2011ib} to the background at the relevant energies. 
The production cross sections for these different processes are summarised in Table\,\ref{nlosigma}.
For the production of $W/Z$+jets we have taken the results provided by {\tt PYTHIA}.
\footnote{Calculations of this quantity at the NLO can be found in e.g., Refs.\,\cite{Berger:2009ep,Melnikov:2009wh}. The uncertainty of the result using {\tt PYTHIA} compared with current data and other simulators can be found in Ref.\,\cite{:2010pg}.}

\begin{table}[h!] \footnotesize
\renewcommand{\arraystretch}{1.50}
\begin{center}
\begin{tabular}{|c||c|c|c|}
\hline 
& 7~GeV & 8~TeV & 14~TeV\\
\hline
$\sigma_{t\bar t}^{NLO}$	&$152_{-19-9}^{+16+8}$~pb 		&$250$~pb $(^*)$				&$852_{-93-33}^{+91+30}$~pb\\
\hline
$\sigma_{WW}^{NLO}$ 	&$47.04_{-3.2\%}^{+4.3\%}$~pb	&$57.25_{-2.8\%}^{+4.1\%}$~pb	&$124.31_{-2.0\%}^{+2.8\%}$~pb\\
$\sigma_{W^+Z}^{NLO}$	&$11.88_{-4.2\%}^{+5.5\%}$~pb	&$14.48_{-4.0\%}^{+5.2\%}$~pb	&$31.50_{-3.0\%}^{+3.9\%}$~pb\\
$\sigma_{W^-Z}^{NLO}$ 	&$6.69_{-4.23\%}^{+5.6\%}$~pb	&$8.40_{-4.1\%}^{+5.4\%}$~pb		&$20.32_{-3.1\%}^{+3.9\%}$~pb\\
$\sigma_{ZZ}^{NLO}$ 	&$6.46_{-3.3\%}^{+4.7\%}$~pb		&$7.92_{-3.0\%}^{+4.7\%}$~pb		&$17.72_{-2.5\%}^{+3.5\%}$~pb\\
\hline
$\sigma_{W+jets}^{LO}$	&$1.46\times10^5$~pb			&$1.74\times10^5$~pb			&$3.50\times10^5$~pb\\
$\sigma_{Z+jets}^{LO}$	&$6.76\times10^4$~pb			&$7.98\times10^4$~pb			&$1.57\times10^5$~pb\\
 \hline
\end{tabular} 
\caption{\small  Cross sections for the production of $t\bar t$ \cite{Kidonakis:2010dk} and $WW/ZZ/WZ$ \cite{Campbell:2011bn} pairs, as well as $W/Z$+jets. $(^*)$ Rough estimate obtained from the data of Ref.\,\cite{Kidonakis:2010dk}.}
\label{nlosigma}
\end{center}
\end{table}

The production cross section of Supersymmetric particles has been computed using {\tt Prospino 2.1}Ê\cite{Beenakker:1996ed}, which provides the result at NLO. The leading contributions obviously comes from the production of coloured sparticles, $\tilde g\tilde g,\,\tilde g\tilde q,\,\tilde q\tilde q$. The actual values are a function of the gluino and squark masses and have been calculated for each specific case.

\begin{figure}[t!]
\hspace*{-0.6cm}
\centering
\includegraphics[width=7.7cm, angle=0]{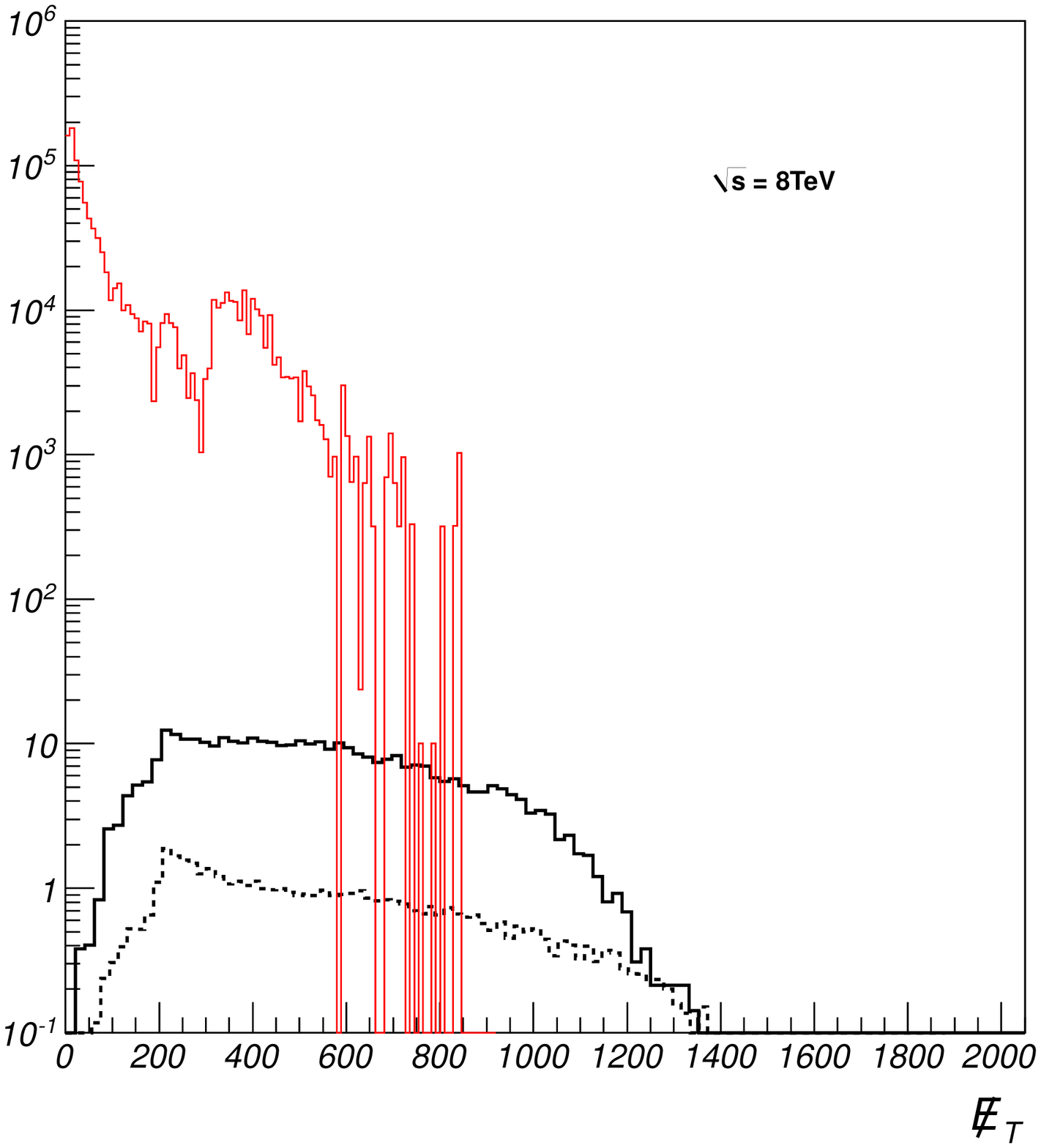}
\includegraphics[width=7.7cm, angle=0]{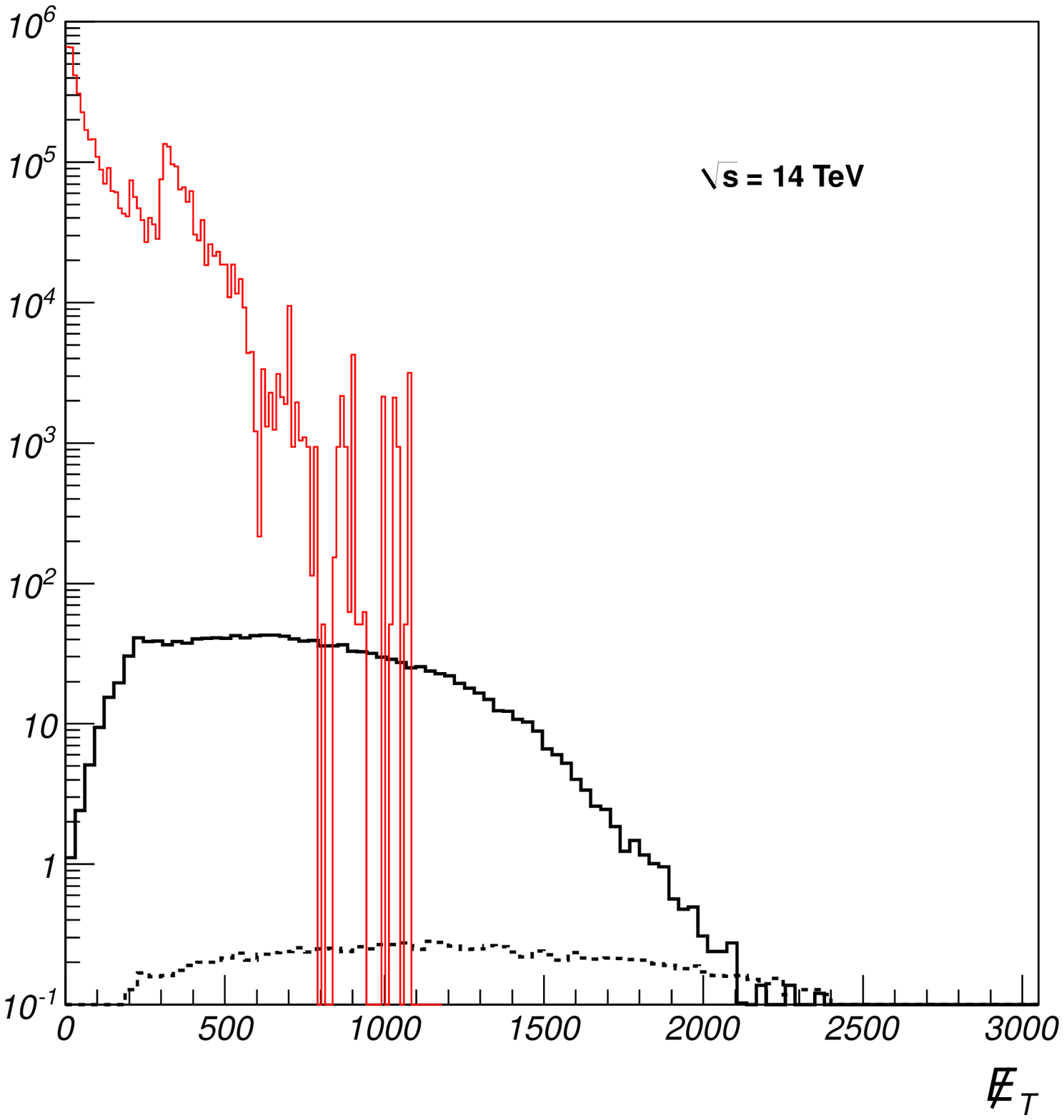}
\caption{Missing energy histogram for the SM background (in red) and SUSY signal in the model. On the left hand-side we assume $\sqrt{s}=8$~TeV and a luminosity of $20$~fb$^{-1}$ and simulate the signal for $M=570$~GeV  (solid line) and $M=700$~GeV (dashed line).
On the right hand-side we assume $\sqrt{s}=14$~TeV and a luminosity of $30$~fb$^{-1}$ and simulate the signal for $M=800$~GeV  (solid line) and $M=1400$~GeV (dashed line).}
\label{plotmet}
\end{figure}

In order to determine the LHC discovery potential we have studied the simplest signal, consisting on the observation of missing transverse energy, $\met$, accompanied by a number ($n\ge 3$) of jets. 
We have used Level 2 triggers in PGS, but supplemented these with additional conditions on the eligible events. Namely, we have implemented the following selection cuts, mimicking those used by the ATLAS Collaboration:
\begin{itemize}\itemsep=0.4ex
 \item[-] Leading jet $P_T$  $>$ 130 GeV,
 \item[-] Second jet $P_T$ $>$ 40 GeV,
 \item[-] Third jet $P_T$ $>$ 40 GeV, 
 \item[-] $m_{eff}$ $>$ 1000 GeV,
\end{itemize}
where $m_{eff}\,=\,E_{T}^{miss} + P_{T}^{jets}$ is calculated from the three leading jets defining the region.
Fig.\,\ref{plotmet} shows a series of histograms for the missing energy resulting from the SM background (red line) and the expected signal events for several examples in the parameter space. In particular, choosing $\sqrt{s}=8$~TeV and a luminosity of 20~fb$^{-1}$ we display the expected signal in our model when $M=570$~GeV and $700$~GeV. Similarly, for  $\sqrt{s}=14$~TeV and a luminosity of 30~fb$^{-1}$ the predictions for the cases $M=800$~GeV and $1400$~GeV are shown.

As we can see, the signal dominates over the background above a given value of the missing energy
with a slight dependence on $M$. The actual number of events obviously depends on the luminosity. 
Given a number of signal events $N_s$ and background events $N_b$ that satisfy our series of cuts, a  statistical
 condition for observability may be defined as
\beq
\frac {N_s}{\sqrt{N_b}}>4 \ ,\  \frac {N_s}{N_b}>0.1 \ ,\ N_s>5\,.
\label{statcondition}
\eeq
It is customary to set a fixed cut for the missing energy in order to determine these numbers, however we have implemented an adaptive method which estimates the optimal value for the cut in $\met$ {\em for each value of $M$}. The idea is to maximize the signal-to-background ratio while guaranteeing that the number of signal events is enough ($N_s>5$).
In particular, if the spectrum is heavy and the signal is expected to be centered around a larger $\met$ then the cut in $\met$ can generally be increased so as to reduce the number of background events as long as the number of signal events is above critical. The latter obviously depends on the luminosity.

\begin{figure}[t!]
\hspace*{-0.6cm}
\centering
\includegraphics[width=7.7cm, angle=0]{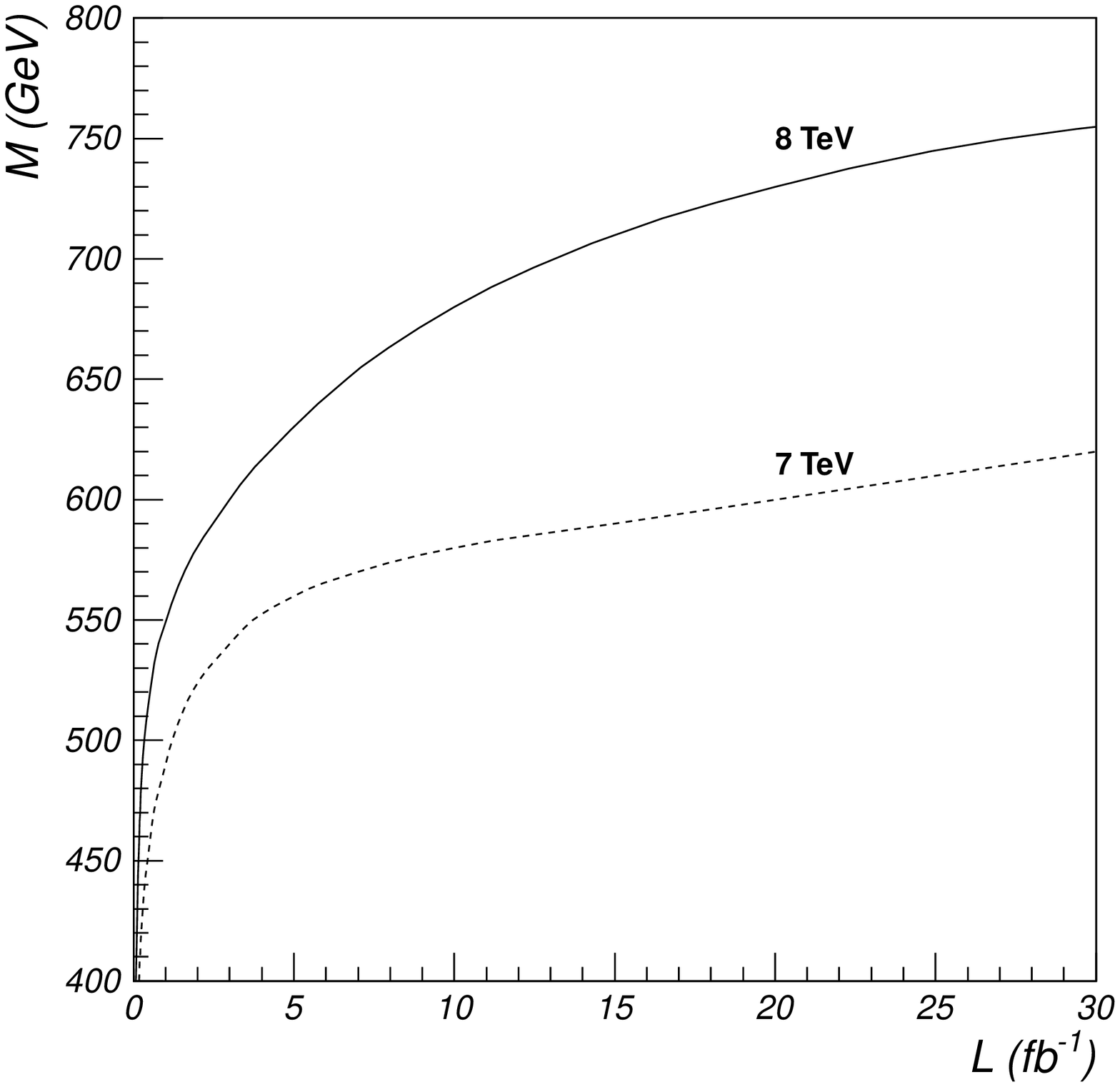}
\includegraphics[width=7.7cm, angle=0]{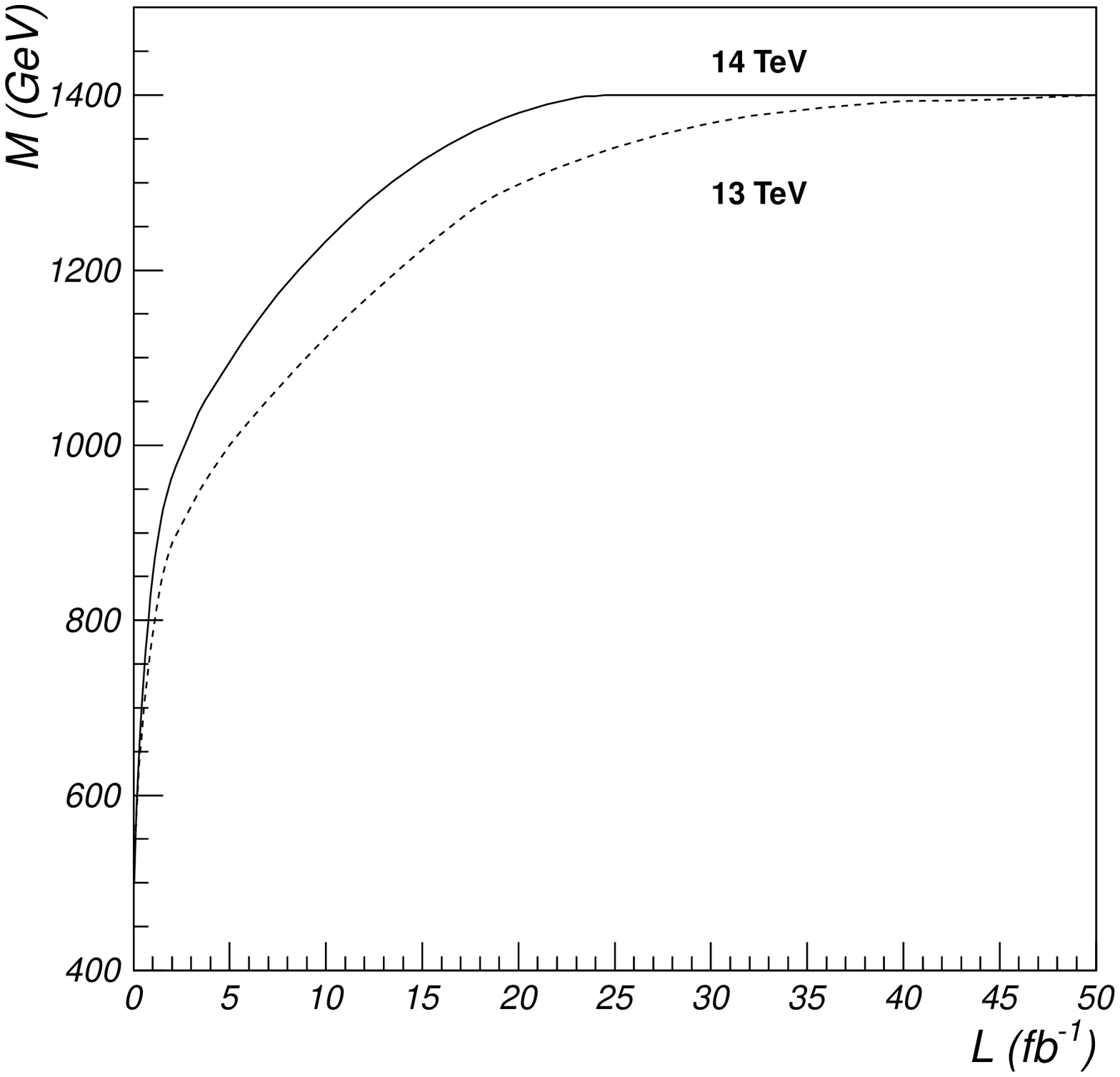}
\caption{Maximum value of $M$ that can be explored at the LHC with $\sqrt{s}=7,\,8$~TeV (left hand-side) and $\sqrt{s}=13,\,14$~TeV (right hand-side) as a function of the luminosity.}
\label{plotluminosity}
\end{figure}

Using this "adaptive cut" in $\met$ we have determined, for each given value of the luminosity (and for each LHC energy configuration), the maximum value of $M$ for which the number of signal events satisfies condition (\ref{statcondition}). This is, we have calculated the detectability potential of LHC for this specific model. 
The results are displayed in Fig.\,\ref{plotluminosity}, where the maximum value of $M$ is plotted as a function of the luminosity. Operating at $\sqrt{s}=7,\,8$ and $13,\,14$~TeV, LHC will be able to test this scenario up to $M\approx  600,\,750$ and $1400$~GeV, respectively, with a luminosity of $20,\,30$ and $30,\,50$~fb$^{-1}$. 
In fact, the LHC at 14~TeV would be able to explore regions of the parameter space with a larger $M$ than the one displayed in the plot. However, as shown in the previous chapters, there is actually no point of the parameter space above that value for which REWSB and dark matter conditions are fulfilled, and for that reason the line flattens at $M=1400$~GeV.

In order to check the validity of our "adaptive cut" in $\met$ we have applied it to the CMSSM and compared the resulting predicted reach with those obtained by the ATLASÊ\cite{atlas} and CMS \cite{cmshadro} collaborations for the same signal. We have obtained a similar reach. Remember in this sense that ATLAS and CMS use a given value for the cut in $\met$ at low masses and a larger value for heavier masses.

\subsection{Other signatures}

As we described in chapter 3, the viable regions of the model correspond to the coannihilation region in which the lightest neutralino and ligtest stau mass are almost degenerate. This class of scenarios has received a lot of attention in the literature \cite{Arnowitt:2001yh,Arnowitt:2006jq,Feldman:2007zn}, since they can give rise to very characteristic signals.
In particular, the following decay chain is dominant for the second-ligest neutralino, $\tilde\chi_2^0\to\tau\tilde\tau_1\to\tau\tau\tilde\chi_1^0$, leading to signals characterised by multiple low energy tau leptons \cite{Arnowitt:2006jq}. In particular, one can search for pairs of opposite sign taus, accompanied by a number of jets, which would be relatively abundant, compared to other characteristic SUSY signals \cite{Feldman:2007zn}.

\subsection{Long-lived staus and Big Bang Nucleosynthesis}

Finally, as we can observe in Fig.\,\ref{plotHiggs1}, the region with larger values of the Higgs mass is precisely that with a smaller mass-splitting between the stau and the lightest neutralino. 
In fact, for Higgs masses above $m_h>124.5$~GeV, for which the recent LHCb constraint on BR($B_s \to \mu^+ \mu^-$) is satisfied,
one finds $m_{\tilde\tau_1}-m_{\tilde\chi_1^0}<1.7$~GeV. This implies that the two body decay $\tilde\tau_1\to\tilde\chi_1^0\tau$ is no longer kinematically allowed and the stau has to undergo three or four body decays ($\tilde\tau_1\to\tilde\chi_1^0\nu_\tau\pi$ or $\tilde\tau_1\to\tilde\chi_1^0\mu\nu_\mu\nu_\tau$). This increases significantly its lifetime which is now larger than $10^{-7}$~s \cite{Jittoh:2005pq}. 
The presence of long-lived staus in the Early Universe has appealing implications for Big Bang Nucleosynthesis (BBN). The stau can form a bound state with nuclei leading to a catalytic enhancement of certain processes (in particular, $^6$Li production) \cite{Pospelov:2006sc}. Moreover, it also provides additional decay processes for $^7$Li and $^7$Be, thereby solving the apparent discrepancy between the observed abundances of these elements and the predicted values in the Standard BBN \cite{Jittoh:2007fr,Jedamzik:2007cp,Pradler:2007is}.  
Indeed, as recently pointed out in Ref.\,\cite{Jittoh:2011ni}, the observed value of $^7$Li can be reproduced if $m_{\tilde\tau_1}-m_{\tilde\chi_1^0}\approx0.1$~GeV, without conflicting with the abundances of the rest of the light elements. 
Remarkably (see Fig.\,\ref{plotHiggs1}), this corresponds to $m_{h}\approx 125$~GeV.

This provides an interesting possibility, the observation of a stable charged particle in the LHC (due to its lifetime, the stau would decay already outside the detector) \cite{Khachatryan:2011ts,Aad:2011hz}.
Notice that staus in these regions have a mass of order 600~GeV, therefore satisfying the
current bounds for long-lived charged particles obtained in ATLAS (at $\sqrt{s} = 7$~TeV and with a luminosity of 37~pb$^{-1}$), which impose $m_{\tilde\tau_1}>135$~GeV at 95\% CL \cite{Aad:2011hz}. 

Finally, long-lived staus might also be searched for in neutrino telescopes after their production inside the Earth from the inelastic scattering of very energetic neutrinos \cite{Ahlers:2006pf,Ahlers:2006pf},Ê although the prediction for their flux is generally very small \cite{Canadas:2008ey}.

\section{A fine-tuned MSSM, a 125 GeV  Higgs  and the landscape}

It is well known that present experimental bounds on SUSY particle masses indicate
a certain amount of fine-tuning at the percent level in the fundamental parameters of
CMSSM models. Our case is no exception, the only difference being that the 
number of fine-tuned parameters is reduced. There are essentially three free
parameters to be tuned: $M$, $\mu$ and $\rho_H$, if we leave coupling constants fixed.
In our scheme there are two independent fine-tunings to be made.
 One is required to get appropriate
REWSB and the other one required for neutralino dark matter, which forces the model
to live in a stau coannihilation region with a good precision. These two conditions leave us
with essentially only one free parameter which may be taken to be the overall scale $M$. 

The question is: why  should nature take those fine-tuned values? In the context of low-energy SUSY 
different approaches have been  followed to understand these fine-tunings (which may be reduced to
only one fine-tuning if one gives up on the dark matter constraint). These range from choosing very particular 
regions of parameter space to reduce fine-tuning or extending the MSSM to include either singlets
(as in the NMSSM) or new gauge interactions. 
Concerning the first possibility,  we have seen that the particular choice of soft {\it and }$\mu$  terms given by
\beq
M\ =\ \sqrt{2}m\ = \ -(2/3)A\ =\ -B \, \  ,  \ \mu=3/2M
\label{magicsoft}
\eeq
gives appropriate REWSB with $M\gg M_Z$ due to delicate cancellations among the different contributions
to $M_Z$. The first set of conditions in Eq.\,(\ref{magicsoft}) are an elegant consequence of modulus dominance
in a large class of models with fermions localized in intersecting 7-branes. 
One could argue there that, if we had a good theoretical reason to hold the  {\it additional}  condition
$\mu=3/2M$, there  would be no fine-tuning. On the other hand getting viable dark matter would require 
$\rho_H\simeq 0.16$, but one could perhaps argue that this  is hardly a  fine-tuning since that is the order
of magnitude of a flux correction. 

\begin{figure}[t!]
\hspace*{-0.6cm}
\centering
\includegraphics[width=12.cm, angle=0]{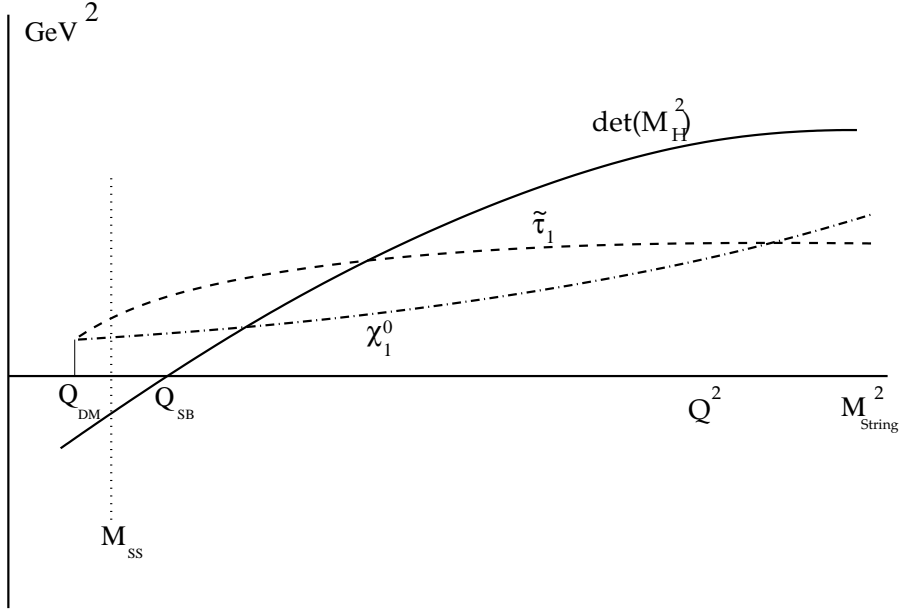}
\caption{The evolution of the Higgs mass$^2$ determinant and ${\tilde \tau}_1$ and $\chi_1^0$ 
mass$^2$ in the infrared.  At the dimensional transmutation scale $Q_{SB}$ a EW Higgs vev 
starts developing. At the scale $Q_{DM}$ the masses of ${\tilde \tau}_1$ and $\chi_1^0$ are degenerate 
signaling the possibility of dark neutralino-stau coannihilation. The SUSY breaking scale $M_{SS}$ is
of order of the  soft SUSY breaking terms and signals the scale where running gets frozen. 
If $M_{SS}$ scans in a landscape one expects $Q_{DM}<M_{SS}<Q_{EWSB}$, with all these
three scales very close in magnitude.}
\label{dimtrans}
\end{figure}

While this reduced fine-tuning would be  tantalizing, there are additional hidden fine-tunings  which 
make this kind of explanation for the {\it little hierarchy} unlikely. In particular, the Higgs potential and sparticle spectrum
depends strongly  on the boundary conditions for the Yukawa couplings of the third generation quarks and leptons. 
Slight variations of these couplings as well as the gauge couplings affect strongly the low energy physics and again 
some fine-tuning of these parameters would have to be made to get REWSB at the right scale.
In addition in the REWSB mechanism there are two important mass scales,  that of SUSY masses $M_{SS}$ and the 
dimensional transmutation scale $Q_{SB}$ at which the Higgs  mass matrix squared starts getting a negative
eigenvalue, see Fig.\,\ref{dimtrans}. Correct REWSB is obtained for $M_{SS}$ slightly below $Q_{SB}$ but those scales
are very sensitive to slight variations of  third generation Yukawa couplings and hence  some fine-tuning 
is again implied.

An alternative is to stick to the MSSM structure and admit that indeed the fine-tuning is there,
in the same way that other relatively small fine-tunings exist in other parameters of fundamental physics. 
One example of this is  the masses of the lightest generation of quarks and leptons, which have
to be in the appropriate ratios so that both the proton is sufficiently stable (so that stable Hydrogen can form) 
and the Deuteron and heavier nuclei are also stable  (see e.g. Refs.\,\cite{antronucl}
for a more detailed discussion).
In this  nuclear stability case  there does not seem to exist a fundamental reason for the ordering and size  of the masses other than
the cosmological
development of appropriate chemical elements which may form the observed world (and us within 
this Universe).  The fine-tuning required on the ratios of Yukawa couplings is in this case
of order $10^{-3}-10^{-2}$.
This suggests an environmental (or anthropic) explanation for the structure of the masses of
the lightest fermion generation, to some extent analogous to  Weinberg's  prediction 
 of a non-vanishing cosmological constant \cite{Weinberg:1987dv} using anthropic arguments.

It may be argued  that the {\it little hierarchy } or fine-tuning problem of the MSSM may be another example
of environmental fine-tuning    \cite{Giudice:2006sn}  analogous to the above mentioned nuclear stability bounds
\footnote{Alternatively it could be that the full weak scale-Planck scale hierarchy
could have an environmental origin, see
Refs.\,\cite{ArkaniHamed:2004fb,Giudice:2004tc,Hall:2009nd,Elor:2009jp}. In these models, though, the Higgs mass tends to
 be heavier than 130 GeV \cite{Cabrera:2011bi,Giudice:2011cg,Arbey:2011ab}.}.
Consider for simplicity of exposition the case with very large $\tan\beta$ in which radiative EWSB is essentially generated
by the Higgs parameter $m_{H_u}^2$ becoming negative in the infrared.  While running down in energies from
a large unification (or string) scale $M_s$ , solving the RGE one gets an expression 
in terms of an adimensional  function $F$ of the form 
\beq
m_{H_u}^2 \ =\ M^2\ F((Q/M_s)^2; \eta, \rho_H) \ ,
\label{trans1}
\eeq
where $M$ is the universal gaugino mass, $\eta=\mu/M$ and we have not displayed additional dependence on gauge and Yukawa coupling constants.
As we said there  is a dimensional transmutation scale $Q_{SB}$ at which  $F(Q_{SB}^2)=0$ and a vacuum expectation value (vev) for the Higgs starts
developing (see Fig.\,\ref{dimtrans}). There is in addition a second independent quantity  $M_{SS}$ which sets the scale of 
SUSY breaking soft terms and sparticle masses. In our scheme $M_{SS}$ is  determined by the RG running of the 
underlying soft terms, which are essentially determined by $M$ and $\mu$. 
For $M_{SS}>Q_{SB}$ the RGE get frozen at a scale of order $M_{SS}$, 
 before $m_{H_u}^2$ becomes negative and symmetry breaking takes place. However,  a universe with unbroken
 electroweak symmetry would be unable to yield a sufficiently complex chemistry for life to develop and hence
 would be untenable on anthropic grounds. On the contrary, for $M_{SS}<Q_{SB}$ the RGE get frozen below the scale $Q_{SB}$,  
 and  $m_{H_u}^2$ gets fixed and negative, yielding EWSB.  It may be argued 
 \cite{Giudice:2006sn}  (see also \cite{Hall:2011jd}) that in a situation in which the soft mass parameter $M_{SS}$ {\it scan} in a landscape 
 of possible values, the most likely situation is one in which $M_{SS}$  is sitting close to $M_{SS}\simeq Q_{SB}$, 
 close to a {\it catastrophic} situation with unbroken EW symmetry.  That precisely corresponds to a 
 a fine-tuned situation with the Higgs vev well below $M_{SS}$ by a one-loop factor \cite{Giudice:2006sn}.

 In our case there is a second relevant dimensional transmutation scale which is close
 to a {\it catastrophic} situation. 
 As we have explained, in our setting one gets appropriate neutralino dark matter only in the
 stau coannihilation region in which  one approximately has $m_{{\tilde \tau }_1}\simeq M_{\chi_1^0}$.
 Outside this region there is a large overabundance of neutralinos (or else the LSP is charged). 
 This is clearly seen in fig.\ref{Mtanbeta}  in which one can observe how correct REWSB {\it and} 
 viable dark matter is only obtained inside a very narrow region in the $M-\tan\beta$ plane. 
 The mass difference  controlling coannihilation is
 \beq
 m_{{\tilde \tau }_1}^2\ - \ 
 M_{\chi_1^0}^2 \ =\ M^2\ G((Q/M_s)^2; \eta, \rho_H) \ .
 \label{trans2}
 \eeq
 At a  scale $Q_{DM}$ such that $G(Q_{DM}^2)=0$ both masses are equal, so that 
 appropriate amount of dark matter is obtained for $M_{SS}>Q_{DM}$, but very close to
 $M_{SS}=Q_{DM}$.  One
 may use again environmental reasoning to argue that  if the  scale $M$ (and hence $M_{SS}$) scans, 
 values of $M$ close to its maximum would be more likely. The absolute environmental maximum would be that
 corresponding to a critical density $\Omega_{matter}=1$ leading to $M\simeq 2.5$ TeV and $m_{h}\simeq 128$ GeV (see Fig.\,\ref{plotHiggs1}).
 On the other hand 
  we saw in chapter 3 that the maximum value consistent with WMAP observations occurs for 
 $M\simeq 1.4$ TeV.
 These  values would   correspond to  the scale $Q_{DM}$ at which ${\tilde \tau}_1$ and $\chi_1^0$ masses are
 approximately equal.  
 All in all we would  have a compressed hierarchy of scales with
 \beq
 Q_{DM}\ \leq \ M_{SS}\leq Q_{SB} \ , 
 \eeq
  in which  environmental criteria would show
 a preference for
 $M_{SS}\simeq Q_{DM}\simeq Q_{SB}$ corresponding to the maximum $M$ compatible with both
 correct REWSB and viable dark matter. This would correspond to the largest Higgs mass values,
 $m_h\simeq 128$ GeV in the extreme case with $\Omega_{matter}=1$ or rather $m_h\simeq 125$ GeV
 if we impose the stronger WMAP bounds (see Fig.\,\ref{plotHiggs1}).
So one can conclude that within the range of Higgs mass values 
 $119$~GeV $\leq m_h\leq 125$~GeV appearing in the present scheme, {\it environmental arguments would
 favour the region close to 125~GeV}.  This is of course essentially a qualitative statement and a more detailed 
 understanding of the different environmental factors playing a role in the combined REWSB mechanism and 
 neutralino dark matter would be needed.

 An interesting question is whether the  basic variables $M$, $\mu$ and $\rho_H$ are likely to scan in a landscape of
 string vacua. In the present context the  MSSM gauge group lives in Type IIB D7-branes (or their F-theory 
 generalizations) with quarks and leptons residing at the intersection of the branes (or matter curves in the F-theory jargon).
 As we described in chapter 2, in the presence of {\it closed string}  ISD backgrounds  soft terms are  generically induced.
 Parameters such as the gaugino mass $M$ correspond to the closed string  flux {\it density} through the branes. On the other hand the 
 {\it open string} magnetic backgrounds through the branes are at the origin of the small correction parameter $\rho_H$.  The origin  of the $\mu$ parameter
 is more model dependent, although indeed closed string fluxes do induce $\mu$-terms in some classes of brane configurations.
 Note that  $M$ and $\rho_H$ are given by {\it local flux  densities},  which are not themselves quantized. Still in a fully fledged Type IIB
 compactification with multiple fluxes those local densities will scan as one varies the possible choices of closed string
 quantized fluxes.  Thus indeed it is reasonable to expect that soft terms do scan in the landscape of Type
 IIB compactifications with fluxes.
 
 In the above argumentation we have ignored 
  that in Eqs.\,(\ref{trans1}) and (\ref{trans2}) there is additional dependence of $F$ and $G$ on
 the gauge and Yukawa couplings. Concerning the gauge couplings they are assumed to be unified at the string scale with the
 (inverse) fine structure constant dependent on the value of the local Kahler modulus $T$. In  the corresponding string vacuum the
 value of  $T$ is expected to be dynamically fixed, with its vev depending on the dynamics induced by the different flux values. 
 This means that the unified couplings will possibly scan, although not necessarily in the same way as soft terms, which are typically 
 directly dependent of fluxes.  Finally, in the above two equations there is dependence on the {\it third generation}  Yukawa couplings, 
 mostly $h_t$, $h_b$  for equation (\ref{trans1}) and $h_\tau$ for (\ref{trans2}). In string compactifications of this large class the
 third generation Yukawa couplings are essentially determined again by the relevant local Kahler moduli like $T$ and hence
 are expected to scan like the unified gauge coupling constant 
 \footnote{This is not the case for the Yukawa couplings of the  first two generations which typically arise from
 instanton corrections in which further dependence on fluxes and other moduli may appear.}.
 Summing up, one expects both gauge and third generation Yukawa couplings to scan in a similar way. On the other hand the qualitative 
 arguments above would  not be modified much by this additional scanning.

\section{Conclusions}

In this paper we have studied in detail several phenomenological aspects of the 
modulus dominance SUSY breaking scheme that we introduced in Ref.\,\cite{aci}. 
These models are theoretically well-motivated since they  are obtained from 
the effective action of a large class of string compactifications. These are Type IIB orientifolds with
quarks/leptons localized at intersecting 7-branes or their F-theory generalizations. 
Rather than a single model, our results apply to one of the largest classes of string compactifications which may lead to realistic 
physics, with all moduli fixed.  

The simplest assumption that the auxiliary field of a modulus field is the origin of SUSY-breaking
leads to a very restrictive set of universal soft terms, Eq.\,(\ref{semboundary}). The presence of magnetic fluxes, required by chirality and symmetry breaking, give rise to small corrections which in the simplest case induce  a slight non-universality
in the Higgs mass parameters. Thus the resulting soft terms, Eq.\,(\ref{boundconditionsfinal}), correspond to a  slice of the CMSSM with a slight non-universal deformation in the Higgs sector.
 Interestingly, this set of well-motivated boundary conditions leads  to a number of attractive features:
1) Correct REWSB, 2) Viable neutralino dark matter for tan$\beta\simeq 41$ in the stau
coannihilation region, 3) Automatic large stop mixing due to the built-in  identity $A=-3/\sqrt{2}m\simeq -2 m$
and large tan$\beta$ required by appropriate dark matter. This allows for a relatively heavy lightest Higgs 
with   119~GeV $\leq m_h \leq $ 125~GeV and a not too heavy SUSY spectrum.
This range narrows down to $m_h\approx125$~GeV when the recent constraint on BR$(B_s\to\mu^+\mu^-)$ is included.
 All these features are remarkable since this is not an ad hoc model 
but was introduced well before LHC data arrived.

Fortunately, this model may be tested at LHC.
The  SUSY spectrum  could start being probed by the LHC at 7~TeV with an integrated luminosity of 5~fb$^{-1}$  (8 TeV with 2~fb$^{-1}$) and the whole parameter space 
would be accessible for 14~TeV and 25~fb$^{-1}$. The signatures would  be quite similar to those of a CMSSM model in the
stau coannihilation region, with very characteristic signatures  involving multi-tau events.
 If the hint of a Higgs at 125 GeV is confirmed, the colored sparticles 
will be heavy but still accessible at LHC at 14 TeV.  On the other hand, for Higgs masses above $124.5$~GeV 
one finds  $m_{{\tilde \tau }_1}-m_{\chi_1^0}\leq  1.7$~GeV and the stau
becomes long-lived, with a life-time longer than $10^{-7}$~s, leaving  a distinctive track at the LHC detectors.
Interestingly, if $m_{{\tilde \tau }_1}-m_{\chi_1^0}\simeq 0.1$~GeV the stau has the right properties to trigger catalytic processes in nucleosynthesis, alleviating the problems associated to the Lithium abundance in standard BBN.
Finally,  improved results for  BR$(B_s\to\mu^+\mu^-)$ from LHCb and CMS in the 2012 LHC run can directly put to test our scheme. If no departure
from the SM value is observed it would be ruled out as it stands. Alternatively, one would have to give up
on the neutralino as a stable LSP and allow for R-parity violation or else allow for a fermion flux correction $\rho_f$ in the
initial soft terms Eq.\,(\ref{boundconditionsfinal}).

Although the number of free parameters in the present model is  reduced compared to
the CMSSM, a certain amount of fine-tuning is still required  both to obtain 
correct REWSB and viable neutralino dark matter. Still, the naturally large stop mixing 
makes possible to obtain a Higgs with a somewhat large mass in the range 119~GeV $\leq m_h \leq  125$~GeV
and {\it at the same time}  a squark/gluino spectrum below 3~TeV, accessible at the LHC. 

Concerning the origin of these fine-tunings, the fact that small deviations from the free parameters 
$M$, $\mu$ and $\rho_H$ and the third generation Yukawa couplings drive the theory into catastrophic regions 
with unbroken EW symmetry and/or above critical matter densities,  may suggest an environmental (anthropic) 
origin.  One may argue along the lines of Ref.\,\cite{Giudice:2006sn} that indeed the {\it little hierarchy} 
problem of the MSSM may have an anthropic explanation. We have seen that the requirement of viable neutralino
dark matter could also add arguments in the same direction. One may argue that 
if soft parameters  and third generation Yukawa couplings  {\it scan}  in a landscape, this would tend to favor the largest values of the 
$M$ parameter consistent with both REWSB and neutralino dark matter, which in turn favour a Higgs mass 
of order 125~GeV.  These  arguments are however only qualitative and a more complete understanding of
the interplay between REWSB and dark matter in environmental selection would be needed.

\vspace{2.0cm}

\newpage

\centerline{\bf \large Acknowledgments}

\bigskip

We thank  C. Albajar, A. Casas, J. Cantero, C.B. Park, J. Terr\'on   for useful discussions. 
This work has been partially supported by the grants FPA 2009-09017, FPA 2009-07908, Consolider-CPAN 
(CSD2007-00042) and MultiDark (CSD2009-00064) from the Spanish MICINN, HEPHACOS-S2009/ESP1473 from the C.A. de Madrid and the contract ``UNILHC" PITN-GA-2009-237920 of the European Commission.
D.G.C. is supported by the MICINN Ram\'on y Cajal programme through the grant RYC-2009-05096.

\end{document}